\documentclass[11pt,twocolappendix]{emulateapj3}
\usepackage{amssymb,amsmath}
\usepackage{graphicx}
\usepackage{natbib}
\usepackage{epstopdf}
\usepackage[draft]{hyperref}

\def\p3m{P${}^3$M} 
\def\ap3m{AP${}^3$M} \def\-{{\em{---}}}

\def\tcc{t_{\rm cc}}

\def\gsim{\;\rlap{\lower 2.5pt
\hbox{$\sim$}}\raise 1.5pt\hbox{$>$}\;}
\def\lsim{\;\rlap{\lower 2.5pt
\hbox{$\sim$}}\raise 1.5pt\hbox{$<$}\;}

\newcommand{\bfB}{\mathbf{B}}
\newcommand{\be}{\begin{equation}}  \newcommand{\ba}{\begin{eqnarray}}
\newcommand{\ee}{\end{equation}}  \newcommand{\ea}{\end{eqnarray}}

 \newcommand{\bi}{\begin{itemize}}
\newcommand{\ei}{\end{itemize}}

\def\lesssim{\mathrel{\hbox{\rlap{\hbox{\lower4pt\hbox{$\sim$}}}\hbox{$<$}}}}
\def\gtrsim{\mathrel{\hbox{\rlap{\hbox{\lower4pt\hbox{$\sim$}}}\hbox{$>$}}}}

\def\pc{p_{\rm CR}}
\def\ec{e_{\rm CR}}
\def\tcc{t_{\rm cc}}
\def\ec{\epsilon_{\rm CR}}
\def\gc{\gamma_{\rm CR}}

\begin{document}

\title{The Launching of Cold Clouds by Galaxy Outflows. IV. Cosmic-Ray-Driven Acceleration.}


\author{Marcus Br\"uggen}
\affil{Universit\"at Hamburg, Hamburger Sternwarte, Gojenbergsweg 112, 21029, Hamburg, Germany}
\author{Evan Scannapieco}
\affil{School of Earth and Space Exploration,  Arizona State University, P.O. Box 871404, Tempe, AZ, 85287-1404}

\begin{abstract}

We carry out a suite of simulations of the evolution of cosmic-ray (CR) driven, radiatively-cooled cold clouds embedded in hot material, as found in galactic outflows. In such interactions, CRs stream towards the cloud at the Alfv\'en speed, which decreases dramatically at the cloud boundary, leading to a bottleneck in which pressure builds up in front of the cloud. At the same time, CRs stream along the sides of the cloud, forming a boundary layer where large filaments develop.  Shear in this boundary layer is the primary mode of cloud destruction, which is relatively slow in all cases, but slowest in the cases with the lowest Alfv\'en speeds. Thus the CR ray pressure in the bottleneck region has sufficient time to accelerate the cold clouds efficiently. Furthermore, radiative cooling has relatively little impact on these interactions.  Our simulations are two-dimensional and limited by a simplified treatment of cosmic-ray dynamics, the neglect of CR heating, and an idealized magnetic field geometry. Nevertheless, our results suggest that cosmic rays, when acting as the primary source of momentum input, are capable of accelerating clouds to velocities comparable to those observed in galaxy outflows. 

\end{abstract}

\section{Introduction}

Galactic winds play a key role in the cosmic cycle of matter,  driving material from deep within the densest regions of galaxies into the rarified fringes of the circumgalactic medium \cite[e.g.][]{2011MNRAS.415...11D,2017ARA&A..55..389T}. Such winds have been studied in great detail across a wide range of galaxy masses and redshifts \citep[e.g.][]{1999ApJ...513..156M, 2001ApJ...554..981P, 2002ASPC..254..292H, 2005ARA&A..43..769V}, including the ultraluminous infrared galaxies (ULIRGs) that provide as much as half of the $z \geq 1$ star formation in the Universe \citep{2005ApJS..160..115R}.  Observations of galactic winds have revealed a complex multi-phase gas ranging from $\approx10^7-10^8$ K plasma observed in X-rays \citep[e.g.][]{1999ApJ...513..156M, 2009ApJ...697.2030S}, to $\approx10^4$  K material observed at optical and near UV wavelengths \citep[e.g.][]{2001ApJ...554..981P, 2007ApJ...663L..77T, 2012ApJ...760..127M, 2012ApJS..203....3S}, to $10-10^3$ K molecular gas observed at radio wavelengths \citep[e.g.][]{2002ApJ...580L..21W, 2013Natur.499..450B}. 

The coexistence of these different phases is not easy to explain, particularly in the presence of cold material moving at supersonic velocities.
Incomplete thermalization of supernova ejecta naturally causes outflows to be born as multiphase gas, but the ram pressure acceleration of cold clouds in hot media is extremely inefficient.  Instead, instabilities and evaporation lead to rapid cloud disruption well before the cold gas  is significantly accelerated.   In the definitive study of the non-radiating, hydrodynamic case, \cite{1994ApJ...420..213K} showed that if hot material moves past a cloud at a velocity $v_{\rm hot}$ that is less than the external sound speed, but greater than the cloud's internal sound speed,  it will shred the cloud within $\approx 2-3$ `cloud crushing' times, defined as  $\tcc \equiv \frac{\chi_0^{1/2} R_{\rm cloud}}{v_{\rm hot}},$
where  $R_{\rm cloud}$ is the cloud radius and $v_{\rm hot}/\chi_0^{1/2}$ is the velocity that the resulting shock moves through the cloud, with $\chi_0$ the initial density ratio between the cloud and the surrounding medium.

When radiative cooling is significant, and thermal conduction is minimal, the primary mechanism for cloud disruption is the Kelvin-Helmholz (KH) instability.
This  is suppressed in supersonic flows, leading to a disruption time that scales as $\tcc \sqrt{1+M}$, where $M$ is the Mach number in the ambient medium \citep[][hereafter Paper I]{2015ApJ...805..158S}. However, when thermal conduction is taken into account, an evaporative interface forms, stabilizing the cloud against hydrodynamical instabilities.  In this case, the clouds can maintain coherent structures for a longer timescale, which can be explained by considering the balance between the internal energy that impinges on a spherical cloud and the energy of the evaporating material
\citep[][hereafter Paper II]{2016ApJ...822...31B}.  In Paper II, we also showed that thermal conduction also changes the shock jump conditions ahead of the cloud and pressure from the evaporation off the surface transforms the cloud into a cylindrical shape that leads to a larger surface, which in turn leads to more evaporation \citep[see also][]{2020MNRAS.tmp.2098H}.

While magnetic fields have the potential to suppress the growth of instabilities, in practice, they do little to increase cloud lifetimes \citep[][hereafter Paper III]{2020ApJ...892...59C}.  Magnetic fields aligned with the wind slow the growth of KH modes, but the magnetic pressure also resists compression and the cloud develops a wispy tail, which mixes with the wind on a timescale similar to that in the hydrodynamic case.   On the other hand, magnetic fields transverse to the wind get `draped' around the cloud, compress the cloud along the field direction, and expand it in the perpendicular direction. The result is continuous mass loss, leading to a shorter lifetime than in the hydrodynamic case. To avoid these problems, various authors have studied alternative models for the cold wind phase, such as the condensation of cold clouds out of the hot phase \citep[e.g.][]{1995ApJ...444L..17W,2003ApJ...590..791S,2016MNRAS.455.1830T,2017ApJ...837...28S,2018ApJ...862...56S}, the recondensation of gas in the tails of cold clouds,   \citep[e.g.][]{2018MNRAS.480L.111G,2020MNRAS.492.1970G} and the acceleration of clouds due radiation pressure \citep[e.g.][]{2005ApJ...630..167T,2011ApJ...735...66M,2011MNRAS.417..950H}. 

Another possibility is the acceleration of cold clouds by cosmic rays (CRs).  Such relativistic particles are thought to play a role in launching galactic winds in Milky Way-type galaxies, where, unlike starbursts, the thermal gas alone is not sufficient to launch a wind. While hot, thermally-driven winds are possible, their mass-loading factors are lower than for the slow cosmic ray-driven winds that advect the warm ($10^4$ K) ionized gas \citep{2018MNRAS.479.3042G}. Similarly, radio haloes in galaxies have long shown the presence of CRs and magnetic fields \citep[e.g.][]{1999AJ....117.2102I,2015AJ....150...81W,2019A&A...628L...3H}.

Several theoretical studies have indicated that CR pressure gradients can lead to outflows \citep{1993A&A...269...54B, 2008ApJ...674..258E, 2012A&A...540A..77D}. \cite{2013ApJ...777L..38H}  demonstrated that CRs alone can launch fast, magnetized winds in high surface density disks. \cite{2013ApJ...777L..16B} and \cite{2014MNRAS.437.3312S} simulated the diffusion of CRs out of star-forming regions and find that this can lead to the establishment of stable vertical pressure gradients that are in line with observations. In global magnetohydrodynamic (MHD) simulations of a Milky Way size galaxy, \cite{2017ApJ...834..208R} showed that moderately super-Alfv\'enic CRs propagate out of denser regions along magnetic fields and thereby accelerate more tenuous gas out of the galaxy. Thus, they concluded that CRs can play an important role in reducing galactic star formation rates and launching galactic winds. 

 \cite{2018MNRAS.476.1680S} devised a spherically symmetric thin shell model for CR-driven outflows and showed that CRs are particularly important for driving outflows in low-mass galaxies. Using idealized simulations of the early phases of outflows, \cite{2020arXiv200700696J} showed that in the early stages of galactic outflows CRs do not have any noticeable effect on the mass loading by the outflow and that in the early stages of galactic outflows the dynamical role of CRs is not important.

When Alfv\'en wave damping processes can be neglected, the CR streaming speed is equal to the Alfv\'en velocity, and CRs are locked to the wave frame. The model of CR streaming was further refined in \cite{2019MNRAS.490.1271H} in which the authors devised a more realistic model of turbulent suppression of the streaming speed. They found that turbulent damping leads to larger scale heights of gas and CRs, and as a result the star formation rate increases with the level of turbulence in the ISM.  \cite{2018ApJ...856..112F} showed results from three-dimensional magnetohydrodynamical simulations that include the decoupling of CRs in the cold, neutral interstellar medium. They found that this decoupling leads to higher wind speed and affects wind properties such as density and temperature. 

A  requirement for a successful acceleration of clouds in galactic halos is that the CR pressure gradient should exceed the gravitational potential. \cite{2010ApJ...717....1L} argued that the CR pressure cannot overcome gravity at high gas surface densities provided that CRs interact with the mean density of the ISM, and if pion losses dominate advection/diffusion. \cite{2018MNRAS.474.4073W} have shown that starbursts act as proton calorimeters in which a substantial fraction of the energy in CRs injected via supernovae is lost through hadronic processes.
Another location where CRs may be essential for the dynamics of cold clouds are high-velocity clouds in the CR halo of the Galaxy \citep{Bregman2005}.

While the aforementioned papers study winds on galactic scales, there is little work on the dynamics of individual clouds in a CR-driven outflow. Recently, \cite{Wiener2019} used 2D simulations to show that a Milky Way sized galaxy with a star formation rate of 20 - 30 $M_{\odot}$/yr would generate enough CRs to accelerate clouds to 100 km s$^{-1}$. They also find that cloud acceleration depends almost linearly on the injected CR flux.
 
In this paper, we study the evolution of cold clouds driven by CR pressure gradients in a magnetized medium. In a suite of two-dimensional simulations performed with the FLASH AMR MHD code \citep{Fryxell2000ApJS}, we expose cold clouds to CR gradients that are driven from one side of the computational domain. The CRs stream with Alfv\'en speeds along the magnetic field lines and thus interact with the cold cloud. Here, we neglect the effect of CR diffusion (both, isotropic and anisotropic) as well as CR heating. To focus on the impact of the CRs themselves, we also neglect thermal conduction as well as gravity, which is negligible for the cloud densities and sizes studied here.

The structure of this work is as follows.  In \S2, we discuss the physics of CR pressure build-up and cloud acceleration and describe our numerical methods and setup.  In \S3 we present our results on the mass loss and velocity resolution and asses the impact of radiative cooling. We present our conclusions in \S4.

\section{Simulations of CR-driven Clouds}

\subsection{Methods}

Since the scattering mean free path of individual CRs is smaller than any other relevant scale, we can treat them as a  fluid.
The transport equation for CRs is then given by
\be
\frac{\partial \pc}{\partial t} = (\gc-1)(\mathbf{u}+\mathbf{v_s})\cdot \nabla\pc- \gc \nabla\cdot[\pc(\mathbf{u}+\mathbf{v_s})],
\ee
where $\pc$ is the CR pressure, $\gamma_{\rm CR}=4/3$ is the 
ratio of the specific heats of the
cosmic rays, $u$ is the velocity of the gas, and $\mathbf{v_s}$ is the streaming velocity \citep[e.g]{1984JPlPh..31..275M}.

Throughout this paper, we work with the CR energy per volume, $\ec$ (mass scalar times density) rather than the pressure and we track its evolution using
the equations of fluid motion. For the simulations, we employ the FLASH code, version 4.3 \citep{Fryxell2000ApJS}, and solve the following equations: 
\begin{eqnarray}
\frac{\partial \rho}{\partial t} + \nabla\cdot\left(\rho \mathbf{u}\right)&=&0 ,\\
\rho \left[
\frac{\partial \mathbf{u}}{\partial t} 
+ (\mathbf{u} \cdot \nabla) \mathbf{u} \right] 
&=& 
\frac{1}{4\pi}(\bfB \cdot \nabla ) \bfB
- \nabla p',   \\
\frac{\partial E}{\partial t} 
+ \nabla\cdot\left(E \mathbf{u} \right)  & = & - 
 \nabla\cdot \left( p' \mathbf{u} \right)  + \frac{1}{4\pi}\nabla\cdot \left[ \left(\bfB\cdot\mathbf{u}\right)\bfB \right], \\
\partial_t \bfB  + (\mathbf{u} \cdot \nabla) \bfB & = &( \bfB \cdot \nabla) \mathbf{u} -\bfB(\nabla\cdot\mathbf{u}),  \\
\nabla\cdot\bfB  &=  & 0,\\
\frac{\partial \ec}{\partial t} +\nabla\cdot (\ec \mathbf{u}) &=& -\mathbf{v_s}\cdot \nabla\ec \\
& & -\gamma_{\rm CR} \ec\nabla\cdot [\mathbf{u+v_s}] +\ec\nabla\cdot \mathbf{u}, \nonumber
\label{eqnmhd}
\end{eqnarray}
where $\rho$, $\mathbf{u}$, $p'=p+\pc+ \frac{1}{8\pi} \left|\bfB\right|^2$, $\bfB$, and $E=\rho \epsilon_\mathrm{int} + 
\rho \frac{1}{2} \left|\mathbf{u}\right|^2 +\frac{1}{8\pi}  \left|\bfB\right|^2$ denote density, velocity, pressure (thermal, cosmic ray, and magnetic), 
magnetic field, and total energy density (internal, kinetic, and magnetic). The first term on the right hand 
side of Eq.~(4) accounts for the magnetic tension due to Lorentz forces, and the second term accounts for the
other component of the Lorentz force, the magnetic pressure. 
The advection of the CRs is handled by the hydro routine and the source terms on the right hand side are handled by a heating routine.  Finally, the CRs affect the gas via $F=-\nabla \pc = -(\gamma_{\rm CR} -1) \nabla \ec,$ which for practical reasons we implement in a modified gravity routine. Eq. (8) does not include a heating term caused by the streaming instability \citep{1975MNRAS.172..557S} well below our simulation resolution limits that thermalize their energy instantly in the thermal gas (sometimes also called streaming losses) \citep{Wiener2019}.

We implemented optically-thin cooling, using the tables compiled by \cite{2009MNRAS.393...99W} from the code CLOUDY \citep{1998PASP..110..761F}, assuming solar metallicity.   As in \cite{2010ApJ...718..417G}, sub-cycling was implemented within the cooling routine itself, such that $T$ and $\Lambda(T, Z)$  were recalculated every time 10\% of the thermal energy was radiated away. For the mean mass per particle we took 0.6 $m_p$. Cooling is disabled for temperatures below $T=10^4$ K.

Following, \cite{2009arXiv0909.5426S} we set the streaming velocity to
\be
\mathbf{v_{\rm s}} = -\frac{B}{\sqrt{4\pi\rho}} \tanh(l_R \nabla \pc /\pc)  \frac{(\bfB \cdot \nabla \pc) \mathbf{\hat b}}{|\bfB \cdot \nabla \pc|},
\ee
where $l_R$ is a regularization parameter that we set to $l_R=50$ kpc and $\mathbf{\hat b}$ is the unit vector in the direction of the magnetic field. Unlike in Wiener et al. (2019), we also regularize the divergence of $\pc$ by the same $\tanh$ function. In addition, we require that $\nabla \pc$ exceeds a threshold of $10^{-35}$ erg cm$^{-4}$ for the $x$-component of the Alfv\'en velocity to turn negative. The latter condition suppresses numerical issues where tiny errors related to the computation of spatial derivatives strongly affect the results and can be physically motivated as wave damping. The timestep limitation imposed by the CR streaming is given by \cite{Wiener2017}:
\begin{eqnarray}
dt_{\rm stream} & \approx &  \frac{dx^2}{l_R |v_s|} \\
& = & 10^{11} \left( \frac{dx}{100\,{\rm pc}}\right )^2 \left( \frac{|v_s|}{200\,{\rm km/s}}\right )^{-1} \left( \frac{l_R}{10\,{\rm kpc}}\right )^{-1} \,{\rm s} . \nonumber
\end{eqnarray}
Moreover, in order to guard against small numerical fluctuations that may reverse the streaming direction of CRs, we set the CR pressure to zero wherever the CR pressure is less than 1 per cent of the value maintained at the lower $x$-boundary. 

Finally, we add a switch that allows acceleration by CRs only on cells whose density lies at least 10 percent above the ambient density, which suppresses the acceleration of the ambient medium by streaming against sound waves and weak shocks that develop from reflections and numerical noise.  It is not immediately clear whether or not this assumption increases in accuracy of the simulation, as Alfv\'en velocity fluctuations due to density differences in the hot medium in real galaxies could potentially contribute to the bulk acceleration of the medium.   We note, however, that: (i) this is unlikely to be main mode of acceleration of the hot medium, which has sound speed much higher than the escape velocity of the host galaxy, and hence expands freely into the surrounding environment \citep{1985Natur.317...44C,2013ApJ...763L..31S}; (ii) the density differences in the hot medium are not well modelled by the numerical noise, and so any acceleration due to this effect in the simulations will not match those nature; (iii) the impact of this threshold on our simulations is small as discussed in more detail below.

\begin{table*}[htp]
\caption{Parameters of the simulations: $\chi$ is the density ratio between cloud and the ambient medium, $\beta$ is the ratio of thermal to magnetic pressure, $B$ is the magnetic field strength, $v_A$ the Alfv\'en velocity in the ambient medium, $v_A/c_s$  the ratio of the Alfv\'en velocity to the sound speed in the ambient medium,  $t_{\rm cc}=\chi_0^{1/2} R_{\rm cl}/ c_{\rm s}$ the cloud crushing time with respect to the sound speed, and $q_0=\pc(x=0) \cdot v_A$ is the source strength.}
\begin{center}
\begin{tabular}{|l|c|c|c|c|c|c|c|c|}
\hline
Name & $\chi$ & $\beta$ & $B$ & $v_A$  & $v_A/c_s$ & $t_{\rm cc}$ & $q_0$  & $N_c$ \\
 &  & & [$\mu$G] & [km/s]  &  & [Myr] & [erg cm$^{-2}$ s$^{-1}$] & [cm$^{-2}$] \\\hline
CHI300BETA100 & 300 & 100 & $5.9\times 10^{-2}$  &    28.7   &   0.11  &    6.5 & $4.0 \times 10^{-8}$ & $6.1\times 10^{18}$ \\\hline
CHI300BETA10 & 300 & 10 & $1.8\times 10^{-1}$    &  90.9   &  0.35    &  6.5  & $1.3  \times 10^{-7}$ & $6.1\times 10^{18}$ \\\hline
CHI300BETA3 & 300 & 3 & $3.4\times 10^{-1}$    &  165.9   &  0.63    &  6.5   & $2.3 \times 10^{-7}$ &$6.1\times 10^{18}$ \\\hline
CHI300BETA1  & 300 & 1 & $5.9\times 10^{-1}$    &  287.5   &  1.1    &  6.5   & $4.0 \times 10^{-7}$ &$6.1\times 10^{18}$ \\\hline
CHI100BETA10 & 100 & 10 & $1.8\times 10^{-1}$   &  52.5   &  0.35     &  6.5  & $7.2 \times 10^{-8}$ & $6.1\times 10^{18}$ \\\hline
CHI100BETA3 & 100 & 3 & $3.4\times 10^{-1}$  &  95.8    &  0.63     &  6.5  & $1.3 \times 10^{-7}$ & $6.1\times 10^{18}$ \\\hline
CHI100BETA1 & 100 & 1 & $5.9\times 10^{-1}$    &  165.9    &   1.1    &  6.5  & $2.3 \times 10^{-7}$ & $6.1\times 10^{18}$ \\\hline
CHI30BETA10 & 30 & 10  & $1.8\times 10^{-1}$  &    28.7 &   0.35 &  6.5  & $4.0 \times 10^{-8}$ & $6.1\times 10^{18}$ \\\hline
CHI30BETA3 & 30 & 3  & $3.4\times 10^{-1}$  &    52.5 &   0.63 &    6.5  & $7.2 \times 10^{-7}$ & $6.1\times 10^{18}$ \\\hline
 CHI30BETA1 & 30 & 1  & $5.9\times 10^{-1}$  &  90.9   &   1.1    &  6.5   & $1.3 \times 10^{-7}$ & $6.1\times 10^{18}$ \\\hline
\end{tabular}
\end{center}
\label{parameters}
\end{table*}

\subsection{Run Parameters}

To better quantify the acceleration of dense clouds by CRs, we set up a simple system in 2D with a source term that maintains a constant CR pressure (equal to the thermal pressure) on the $x=0$ boundary. The physical size of the domain was 8 kpc in the $x$-direction and 3.2 kpc in the $y$-direction. The root grid size was 10 $\times$ 4 blocks of $8\times 8$ cells and a maximum of 5 levels of refinements yielding an effective resolution of $\Delta x  = 1.6$ pc  with $5120 \times 2048$ cells. 

At a distance of $325$ pc from the boundary, we placed a circular cloud with a radius of $R=100$ pc,  which is equal to 64 resolution elements.
The density inside the cloud was $\rho_{\rm cl}=10^{-26}$ g cm$^{-3}$ and the ambient density was $\rho_{\rm amb}= \rho_{\rm cl}/\chi$.  In order to avoid numerical artifacts at boundaries with infinite gradients in the Alfv\'en speed, we used Gaussian tapers at the edges of the cloud with scale lengths that equal a fourth of the cloud radius, e.g. $\rho(r)=\rho_{\rm amb}+(\rho_{\rm cl}-\rho_{\rm amb})\exp(-r^2/l_s^2)$, where $r$ is the radial distance from the edge of the cloud and $l_s$ is the scale length. The temperature inside the cloud for all runs was fixed at $10^4$ K and the temperature in the ambient gas was set by demanding that the initial thermal pressure is constant throughout the simulation volume.  The Jeans mass of the cloud after shock passage at a temperature of $10^4$ K and number densities of the order of 10 cm$^{-3}$ are several $10^7M_\odot$, which is more than the cloud masses considered here.

As in our previous papers in the series, we tracked the cloud using a massless scalar and computed its center-of-mass (COM) position ${\bf x_{\rm c}}$ and velocity  ${\bf v_{\rm c}}$, as well as its radial extent in the $x$ and $y$ directions, calculated as the mass-weighted average values of ${\rm abs}(x-x_{\rm c})$ and ${\rm abs}(y-y_{\rm c})$.

Outside the cloud, the magnetic field was assumed to be uniform in the $x$-direction, such that the CRs stream towards the cloud, and the field strength was set by the plasma parameter $\beta=p_{\rm therm}/p_B$. We did not investigate other magnetic field configurations. In our set-up, transverse magnetic fields would not allow the CRs to reach the cloud. In a set of runs we varied the dimensionless parameters $\chi$ and $\beta$ as listed in Tab.~\ref{parameters}. 
 The parameters are chosen such that the Alfv\'en velocity, $v_{\rm A}\sim 15\,{\rm km/s} \sqrt{\chi/\beta} \gtrsim 25 ,{\rm km/s}$, i.e. $\chi/\beta > 3$. 
 The cloud crushing time with respect to the sound speed is $\chi_0^{1/2} R_{\rm cl}/ c_{\rm s},$ which is also given in Tab.~\ref{parameters}. Hereafter, we use this definition of the cloud crushing time.

\begin{figure*}[htbp]
\begin{center}
\includegraphics[trim=0 50 0 50,clip, width=0.24\textwidth]{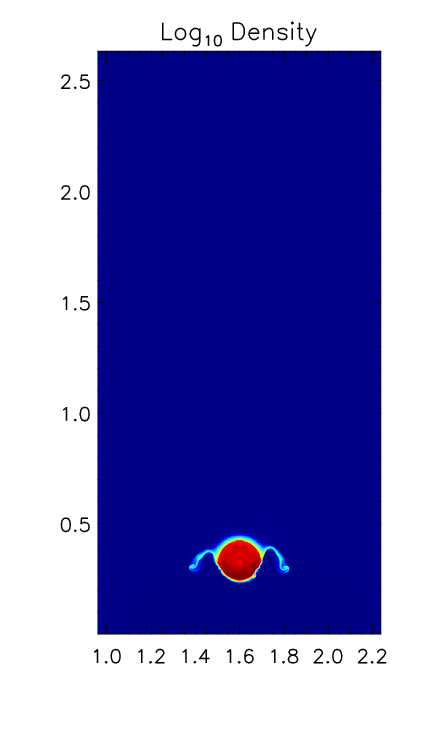}
\includegraphics[trim=0 50 0 50,clip, width=0.24\textwidth]{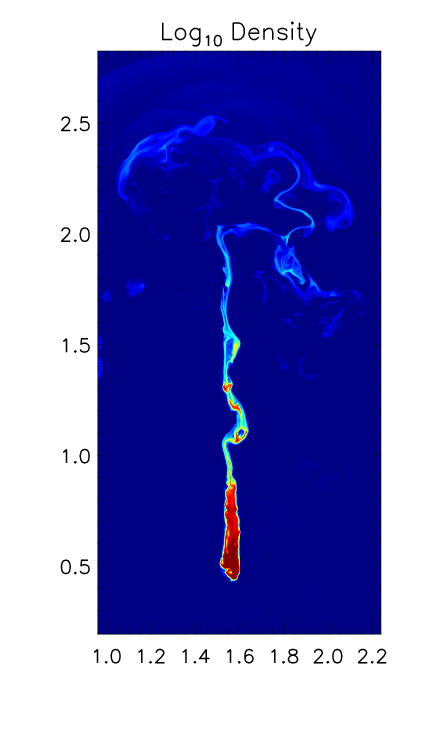}
\includegraphics[trim=0 50 0 50,clip, width=0.304\textwidth]{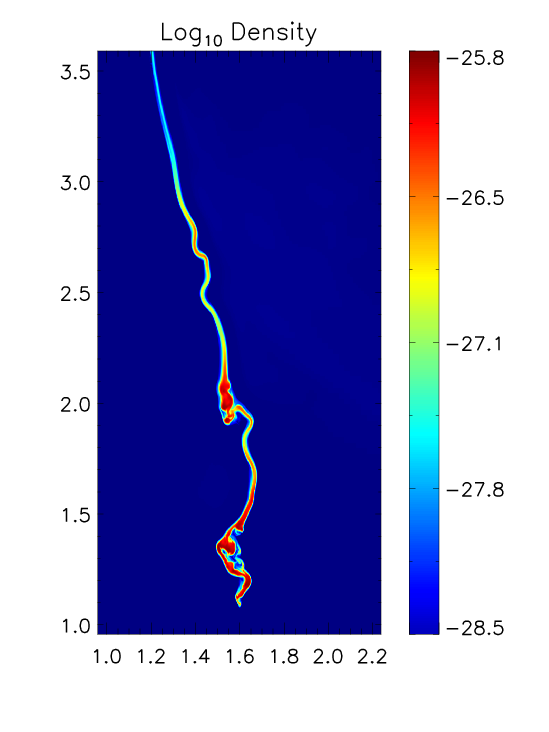}\\
\includegraphics[trim=0 50 0 50,clip, width=0.24\textwidth]{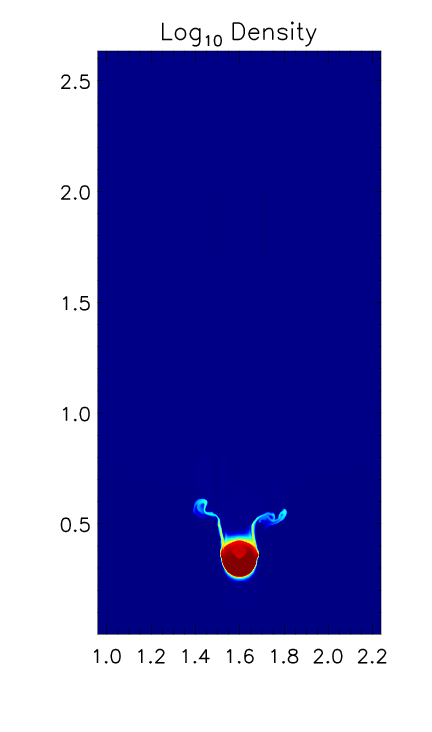}
\includegraphics[trim=0 50 0 50,clip, width=0.24\textwidth]{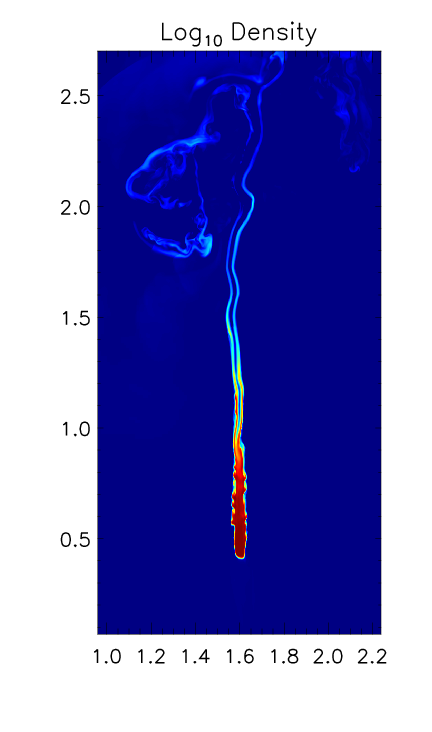}
\includegraphics[trim=0 50 0 50,clip, width=0.304\textwidth]{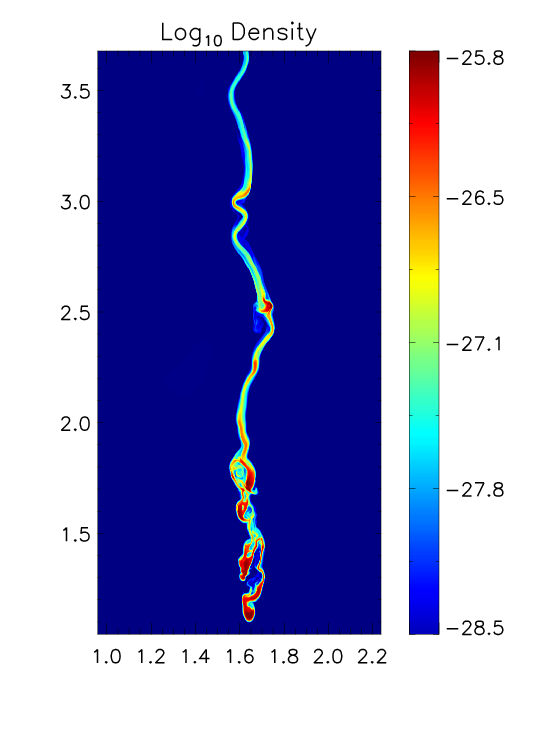}\\
\includegraphics[trim=0 50 0 50,clip, width=0.24\textwidth]{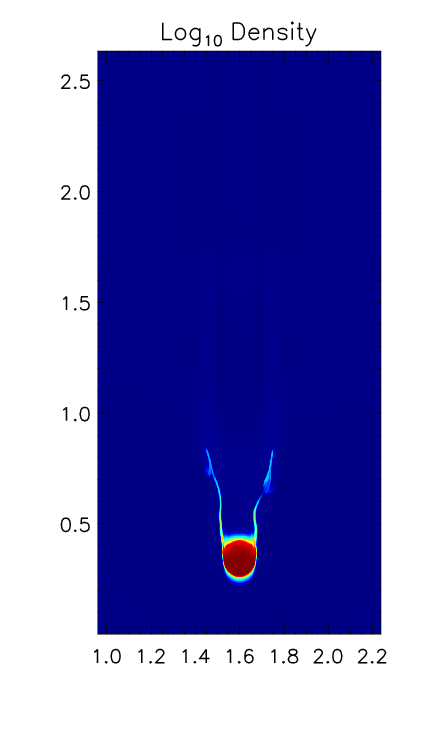}
\includegraphics[trim=0 50 0 50,clip, width=0.24\textwidth]{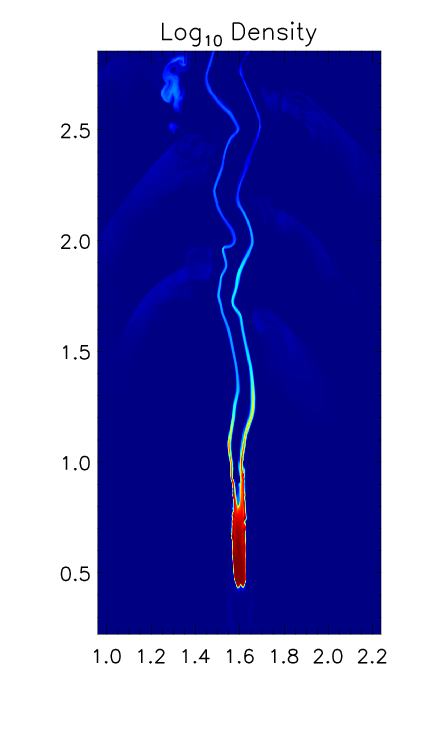}
\includegraphics[trim=0 50 0 50,clip, width=0.304\textwidth]{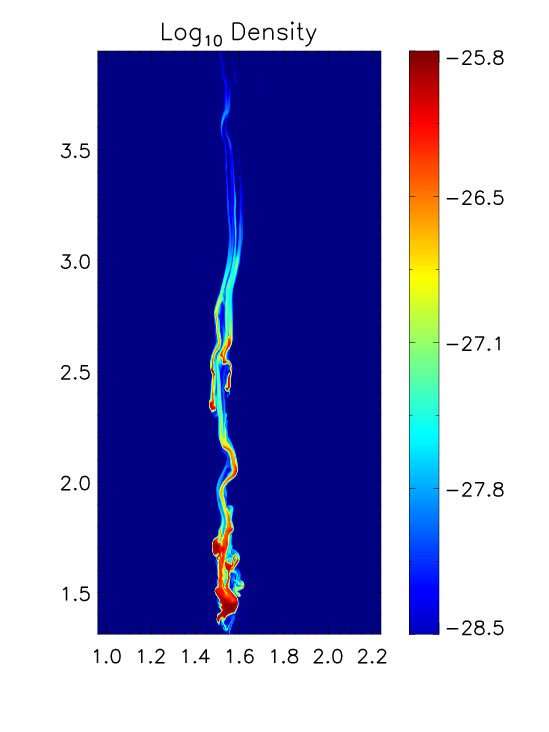}
\caption{Snapshots of gas density at  $t=1$ (left), 5 (center),  and 10 $t_{\rm cc}$ (right).  From top to bottom: CHI300BETA100, CHI300BETA10, CHI300BETA3. Here and in the figures below, the axes labels denote kpc. }
\label{dens300}
\end{center}
\end{figure*}

\begin{figure*}[htbp]
\begin{center}
\includegraphics[trim=0 50 0 50,clip, width=0.24\textwidth]{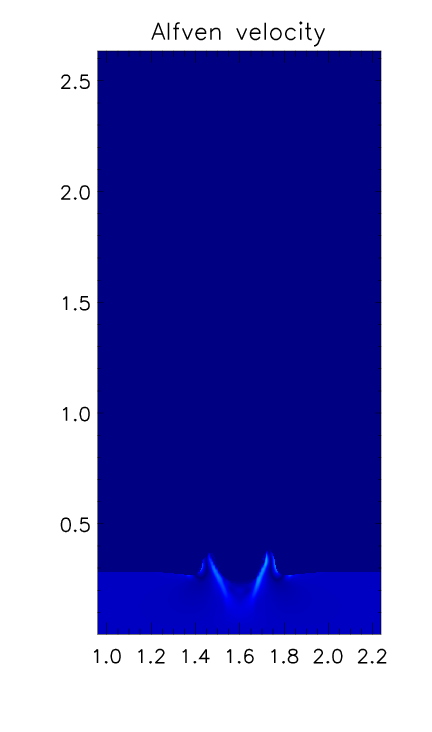}
\includegraphics[trim=0 50 0 50,clip, width=0.24\textwidth]{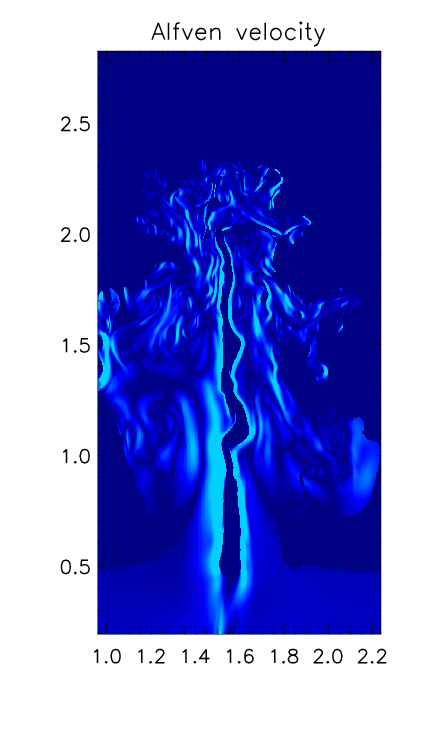}
\includegraphics[trim=0 50 0 50,clip, width=0.304\textwidth]{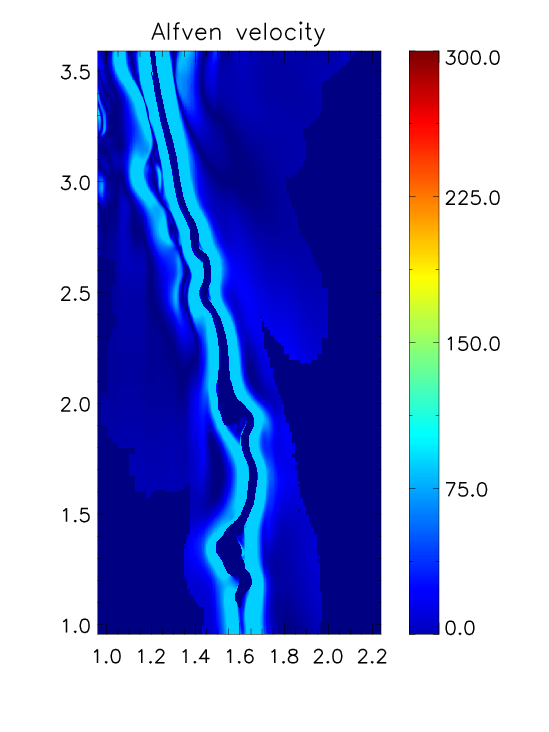}\\
\includegraphics[trim=0 50 0 50,clip, width=0.24\textwidth]{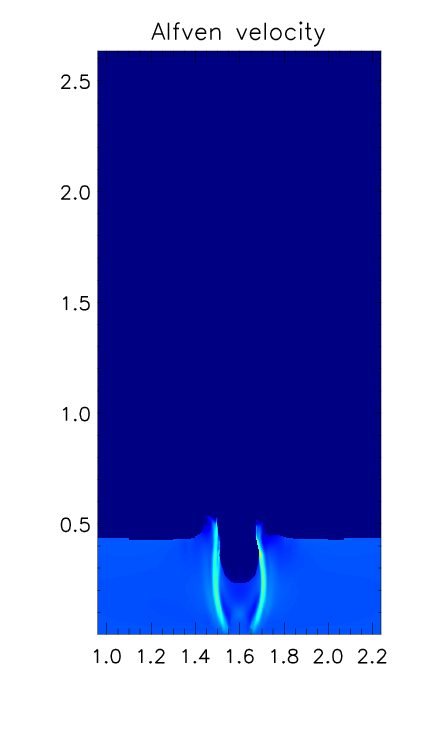}
\includegraphics[trim=0 50 0 50,clip, width=0.24\textwidth]{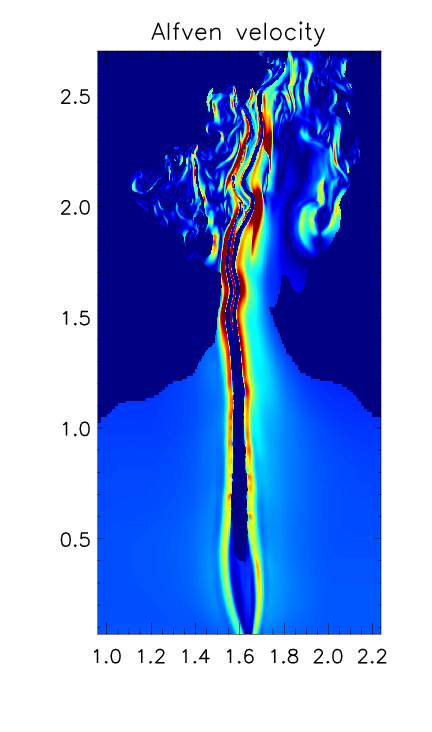}
\includegraphics[trim=0 50 0 50,clip, width=0.304\textwidth]{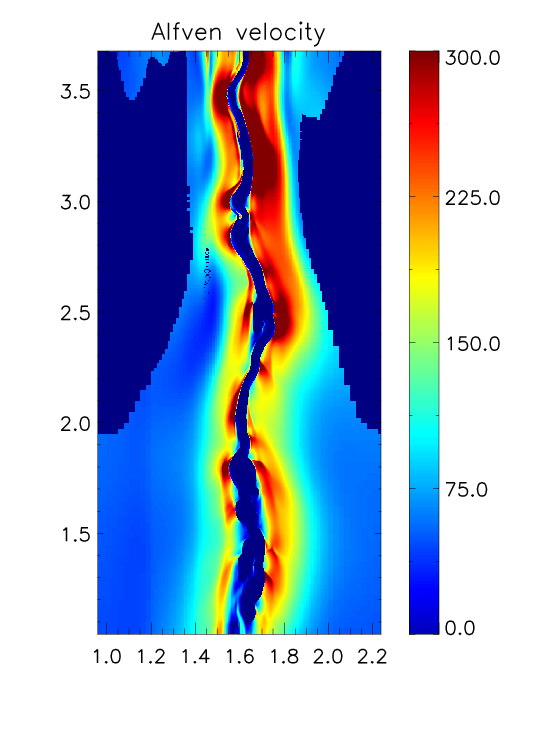}\\
\includegraphics[trim=0 50 0 50,clip, width=0.24\textwidth]{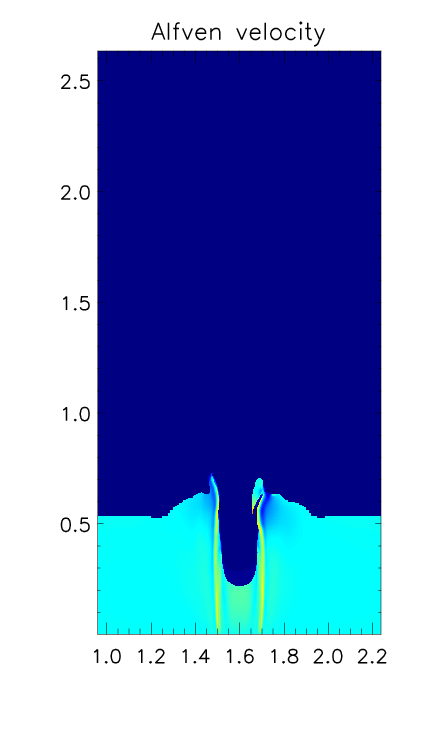}
\includegraphics[trim=0 50 0 50,clip, width=0.24\textwidth]{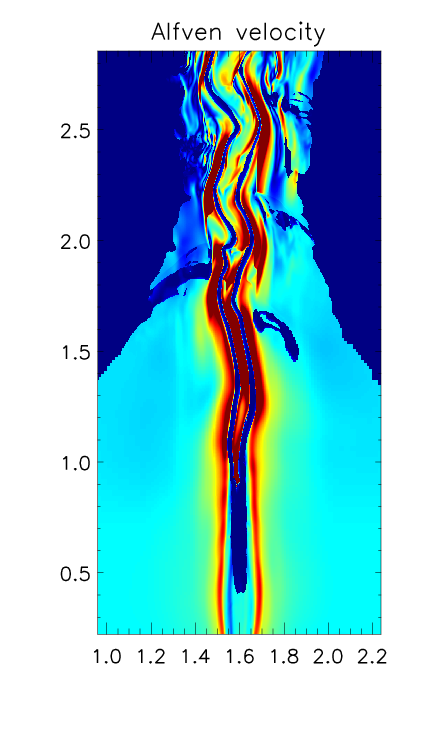}
\includegraphics[trim=0 50 0 50,clip, width=0.304\textwidth]{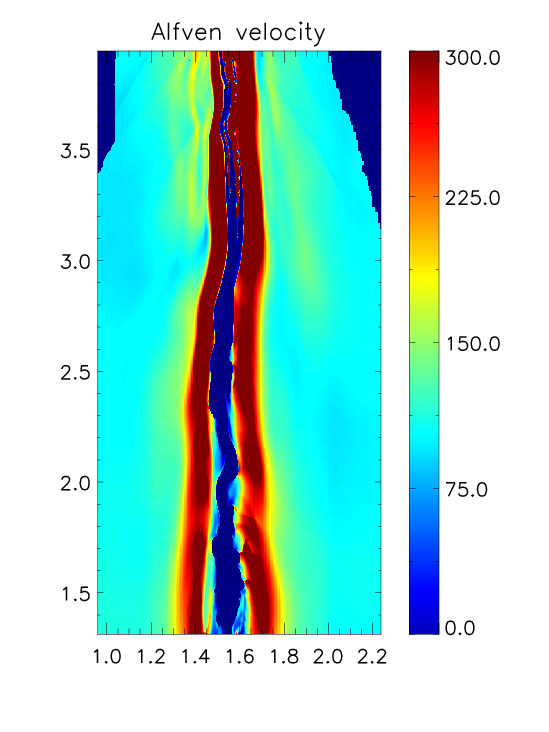}
\caption{Snapshots of Alfv\'en velocity at   $t=1$ (left), 5 (center),  and 10 $t_{\rm cc}$ (right). From top to bottom: CHI300BETA100, CHI300BETA10, CHI300BETA3.}
\label{ax300}
\end{center}
\end{figure*}

\begin{figure*}[htbp]
\begin{center}
\includegraphics[trim=0 50 0 50,clip, width=0.24\textwidth]{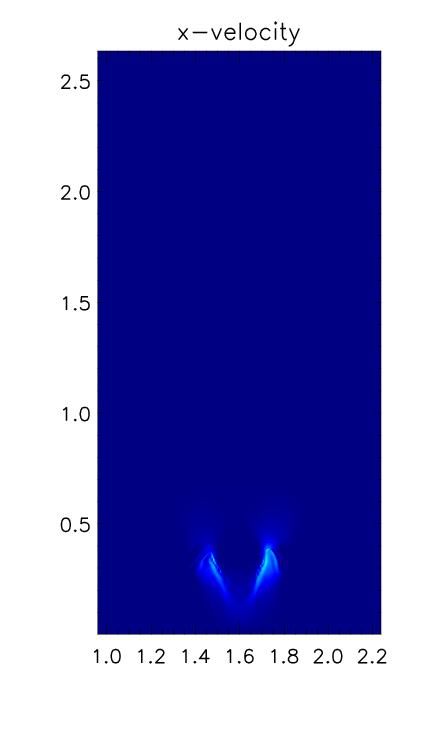}
\includegraphics[trim=0 50 0 50,clip, width=0.24\textwidth]{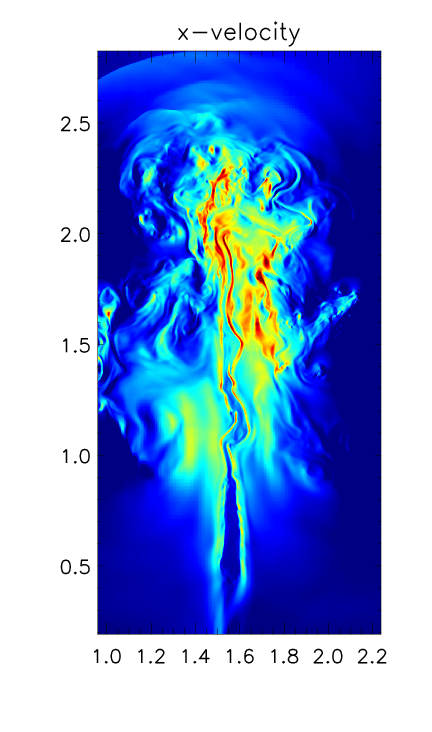}
\includegraphics[trim=0 50 0 50,clip, width=0.304\textwidth]{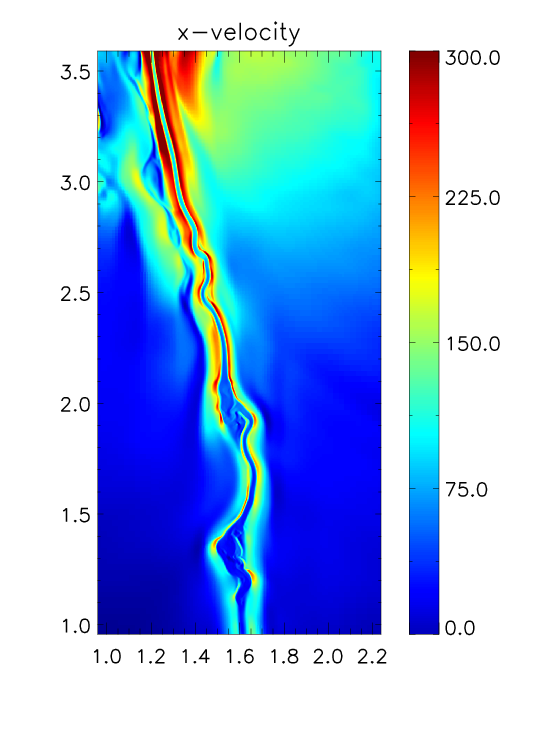}\\
\includegraphics[trim=0 50 0 50,clip, width=0.24\textwidth]{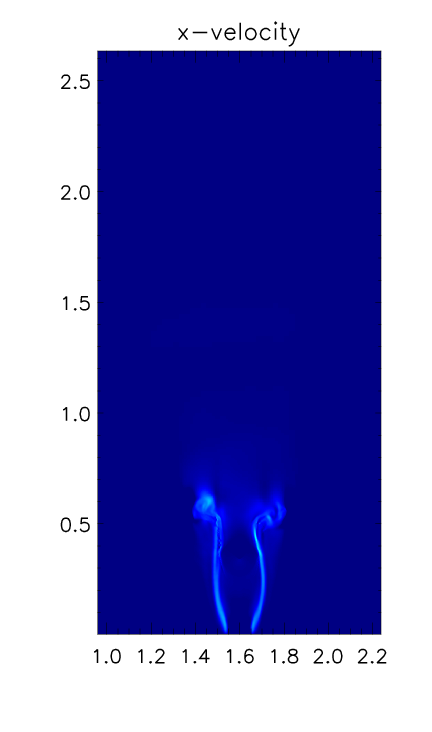}
\includegraphics[trim=0 50 0 50,clip, width=0.24\textwidth]{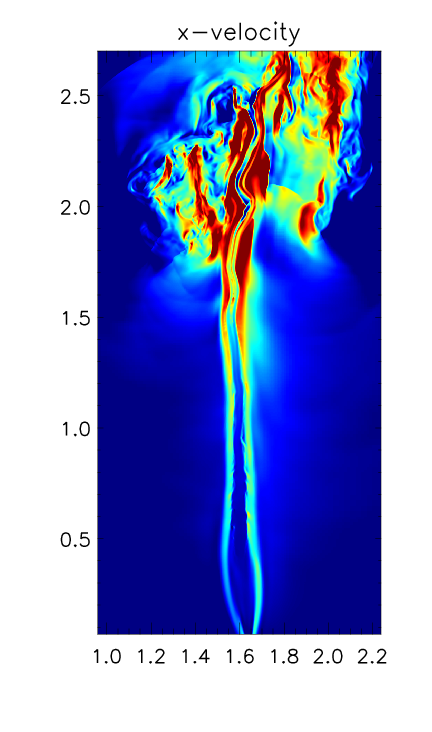}
\includegraphics[trim=0 50 0 50,clip, width=0.304\textwidth]{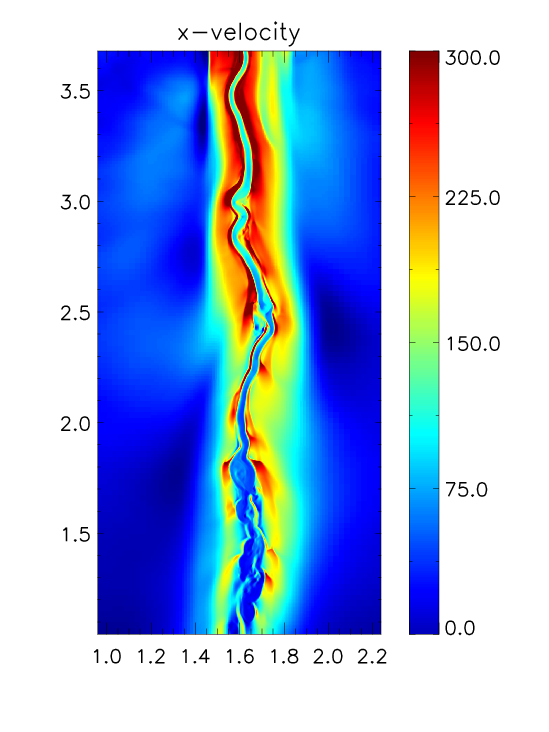}\\
\includegraphics[trim=0 50 0 50,clip, width=0.24\textwidth]{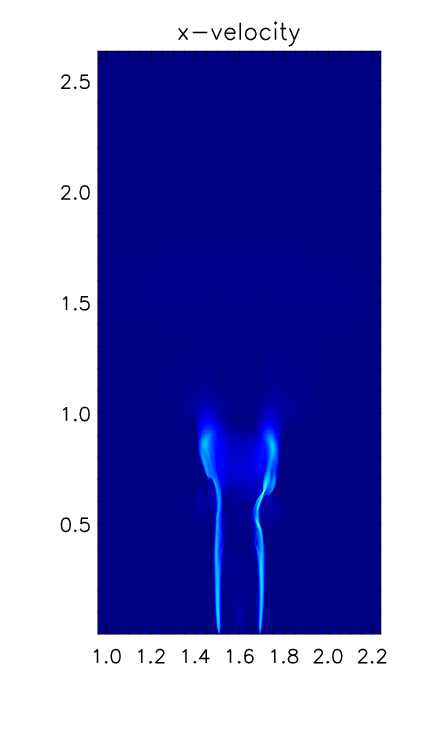}
\includegraphics[trim=0 50 0 50,clip, width=0.24\textwidth]{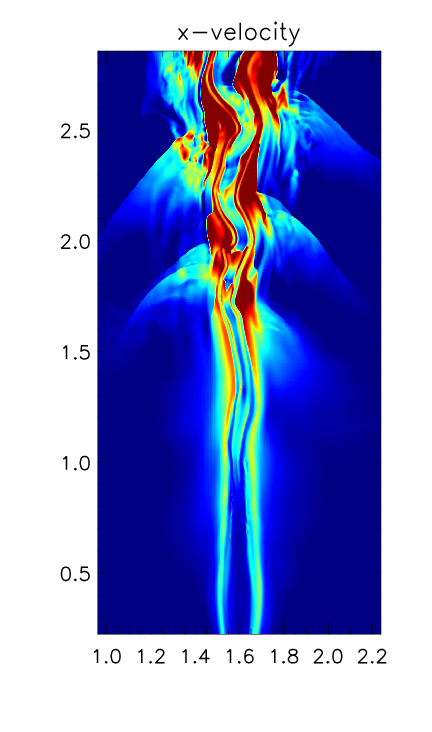}
\includegraphics[trim=0 50 0 50,clip, width=0.304\textwidth]{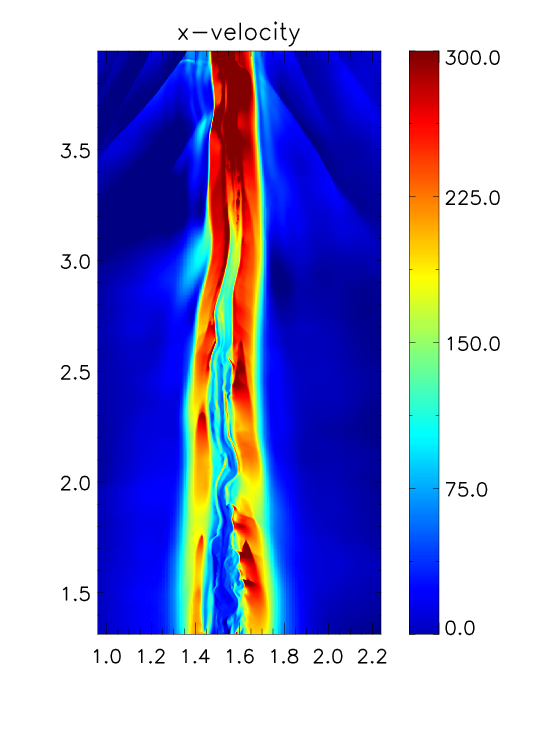}
\caption{Snapshots of gas $x$-velocity at $t=1$ (left), 5 (center),  and 10 $t_{\rm cc}$ (right). From top to bottom: CHI300BETA100, CHI300BETA10, CHI300BETA3.}
\label{v300}
\end{center}
\end{figure*}

\section{Results}

\subsection{Morphological Evolution}

In Fig.~\ref{dens300}, we show the density of our $\chi=300$ runs at fixed cloud crushing times of 1, 5,  and 10 $t_{\rm cc}$, illustrating the overall evolution of the interaction.  The CRs stream towards the cloud at the Alfv\'en speed, which decreases dramatically at the cloud boundary.   This leads to a bottleneck in which the CRs pile up in front of the cloud \cite[e.g.][]{Wiener2017}, building up pressure, which compresses the cloud in the direction of the flow and begins accelerating it downstream.

At the same time, the CRs start to stream along the sides of the cloud, leading to a second pressure gradient.  This causes the cloud to be compressed in the perpendicular direction, dragging the field lines with it and increasing the magnetic field strength.  Note that due to magnetic tension, the field lines that thread the cloud are pushed together not just within the cloud, but also in the regions of the hot medium immediately surrounding it.  This leads to a region of high magnetic field strength, but low mass density in the boundary layer directly surrounding the cloud, causing the Alfv\'en speed to go up dramatically within these areas, as illustrated in Fig.~\ref{ax300}.

As a result, the CRs stream past the cloud, and the pressure gradient associated with the additional pressure significantly increases the downstream velocity of the gas in this layer.  The flow is then divided into three regions, as can be seen in Fig.~\ref{v300}: (i) the bottleneck region immediately in front of the cloud, in which CR pressure causes cloud acceleration but  gas velocities are relatively modest, (ii) a boundary layer on the sides of the cloud, in which the CR pressure compresses the cloud perpendicular to the flow, and gas velocities are large, leading to significant shear, and (iii) the region far away from the cloud, in which CRs stream freely along the magnetic fields and the gas velocities are small. 

Region (i) is most important in the acceleration of the cloud, but region (ii) is the most important in determining its morphology, as well as the cloud lifetime.  This shear layer causes mass to be ablated from the cloud in long, narrow filaments that are advected downstream with the wind for several kpc, and the rate of this ablation is determined by the KH instability in the boundary region. Note that for a given density contrast, the filaments are longer for cases with lower values of $\beta,$ in which the magnetic field plays a larger role in shaping the dynamics of the flow. Lower values of $\beta$ also imply higher Alfv\'en speeds, and hence faster CR streaming.   This is in contrast to the gas velocities, which depend primarily on the CR pressures and are very similar across the different $\chi=300$ runs.

\begin{figure*}[htbp]
\begin{center}
\includegraphics[trim=0 50 0 50,clip, width=0.24\textwidth]{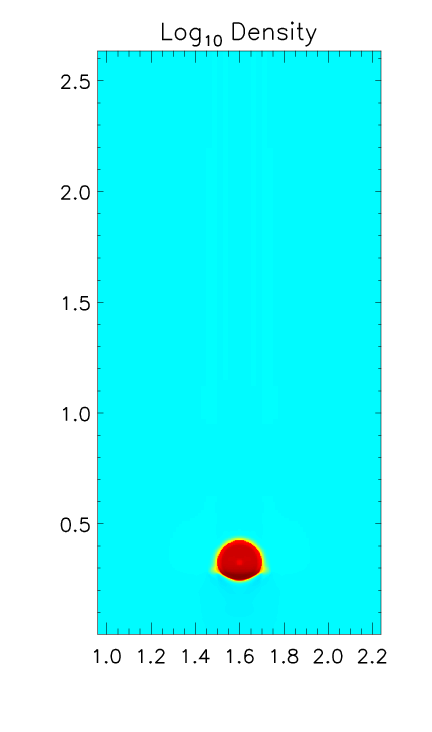}
\includegraphics[trim=0 50 0 50,clip, width=0.24\textwidth]{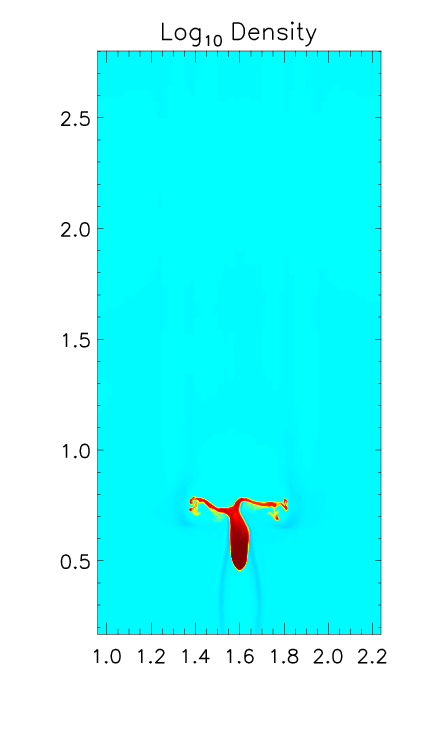}
\includegraphics[trim=0 50 0 50,clip, width=0.304\textwidth]{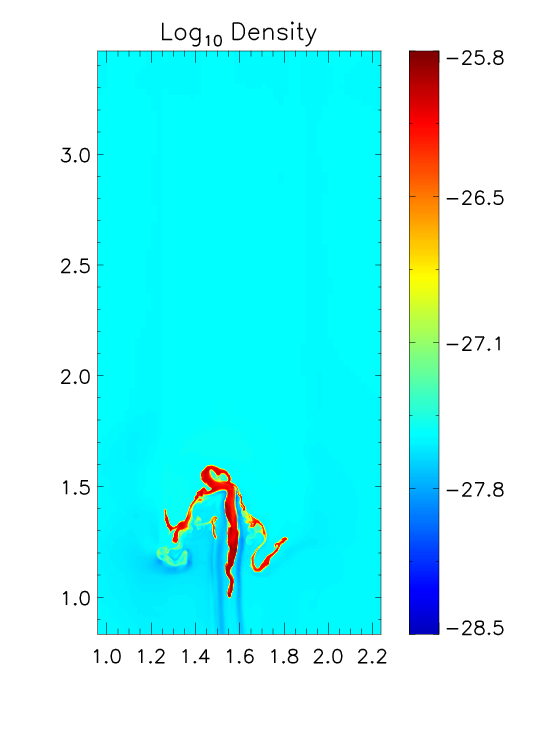}\\
\includegraphics[trim=0 50 0 50,clip, width=0.24\textwidth]{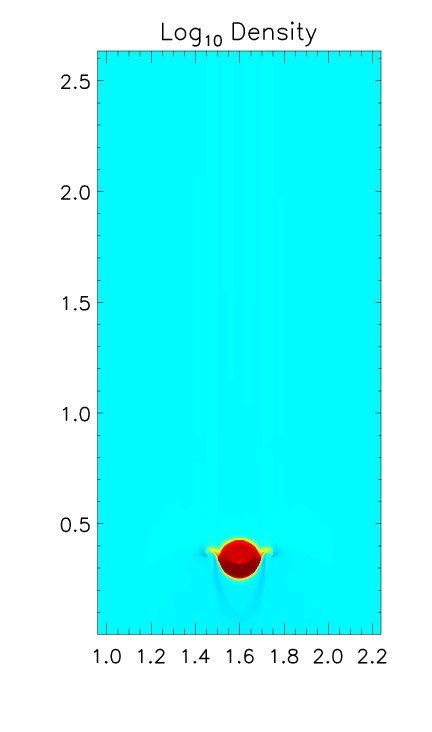}
\includegraphics[trim=0 50 0 50,clip, width=0.24\textwidth]{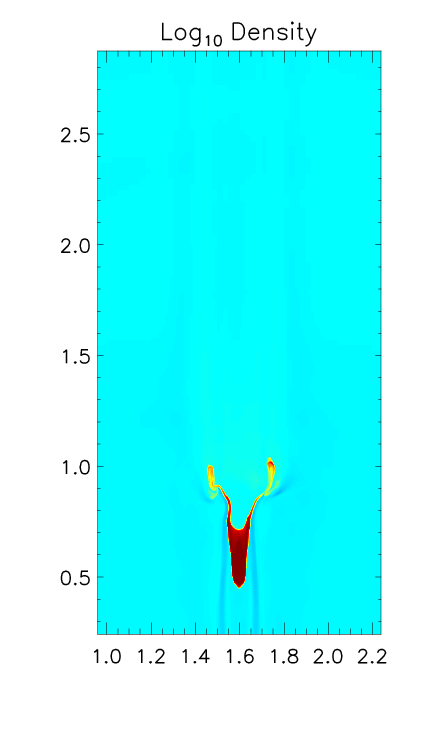}
\includegraphics[trim=0 50 0 50,clip, width=0.304\textwidth]{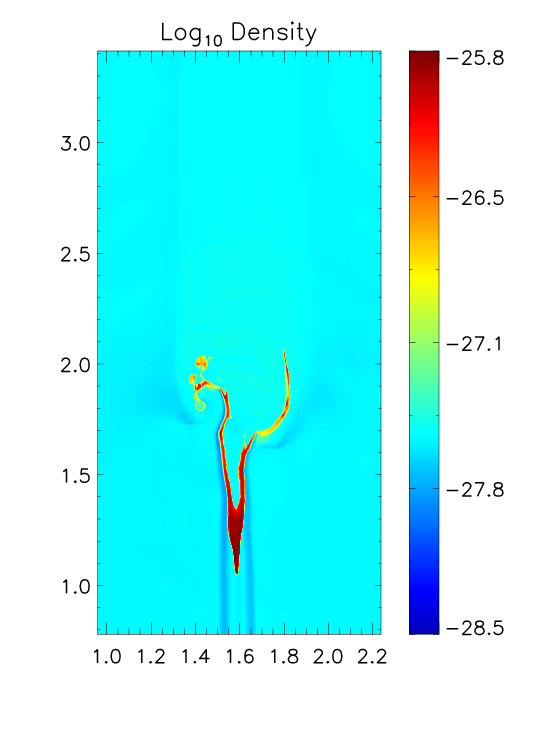}\\
\includegraphics[trim=0 50 0 50,clip, width=0.24\textwidth]{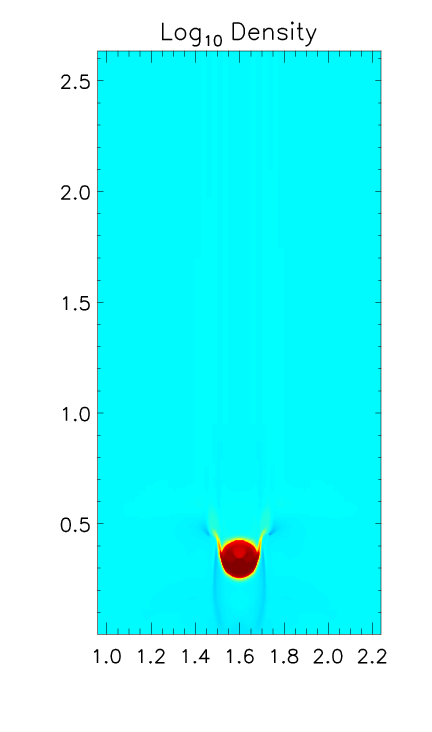}
\includegraphics[trim=0 50 0 50,clip, width=0.24\textwidth]{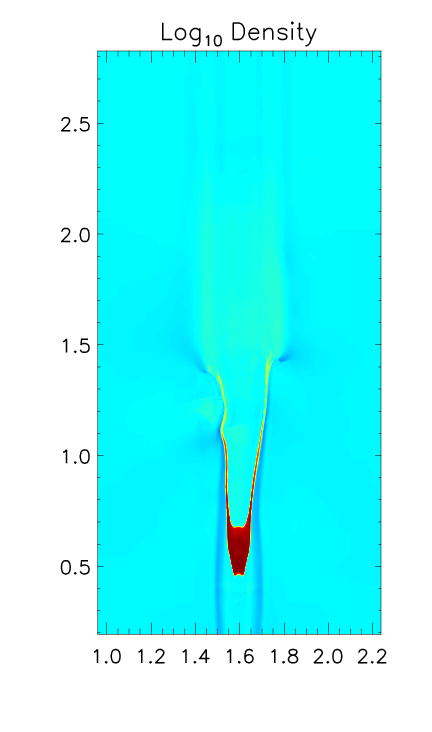}
\includegraphics[trim=0 50 0 50,clip, width=0.304\textwidth]{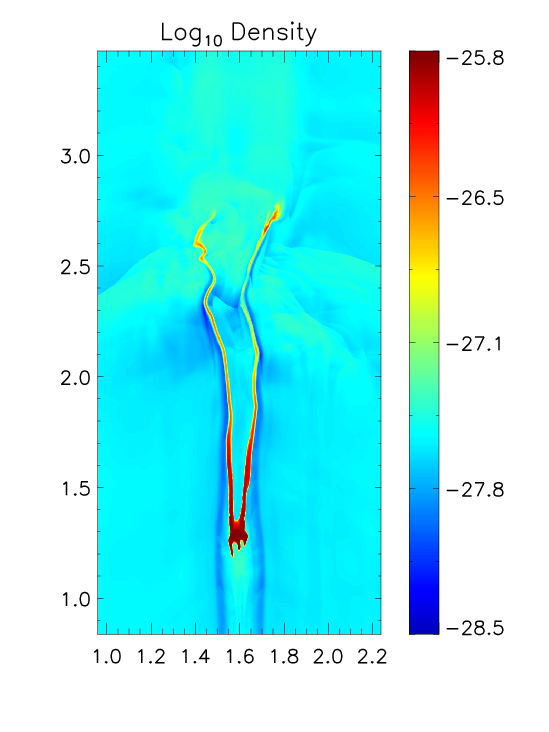}
\caption{Snapshots of gas density at   $t=1$ (left), 5 (center),  and 10 $t_{\rm cc}$ (right).  From top to bottom: CHI30BETA10, CHI30BETA3, CHI30BETA1.}
\label{dens30}
\end{center}
\end{figure*}

\begin{figure*}[htbp]
\begin{center}
\includegraphics[trim=0 50 0 50,clip, width=0.24\textwidth]{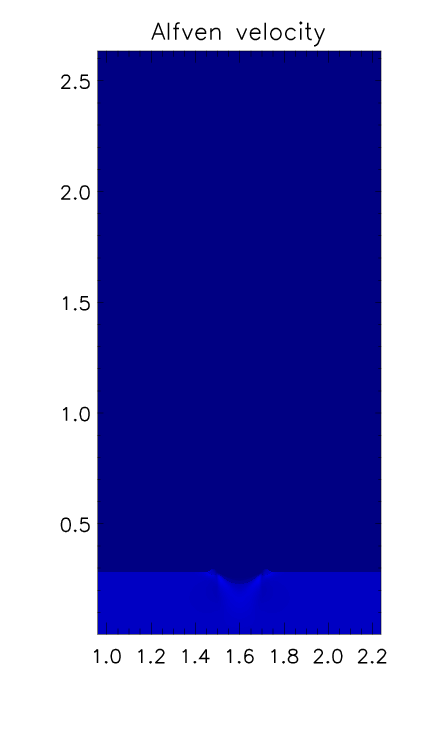}
\includegraphics[trim=0 50 0 50,clip, width=0.24\textwidth]{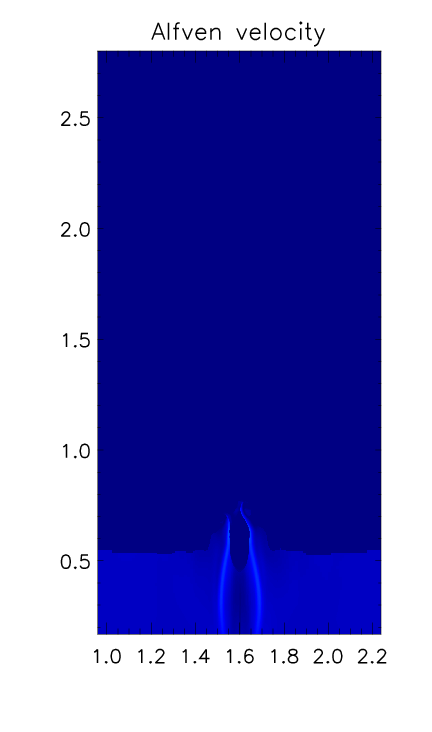}
\includegraphics[trim=0 50 0 50,clip, width=0.304\textwidth]{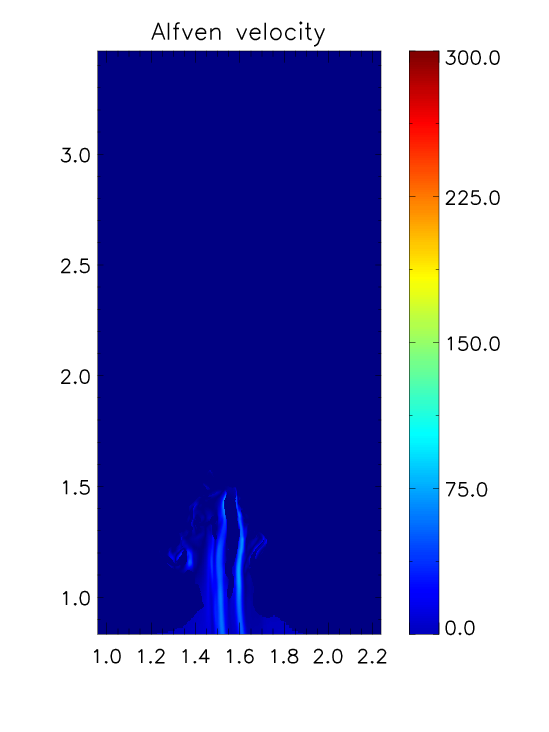}\\
\includegraphics[trim=0 50 0 50,clip, width=0.24\textwidth]{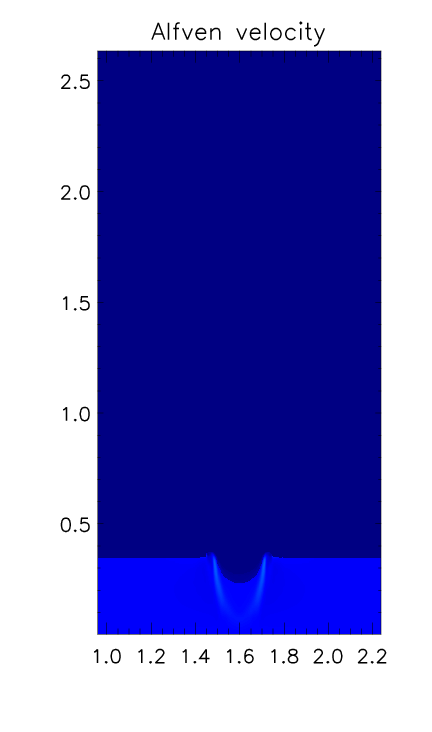}
\includegraphics[trim=0 50 0 50,clip, width=0.24\textwidth]{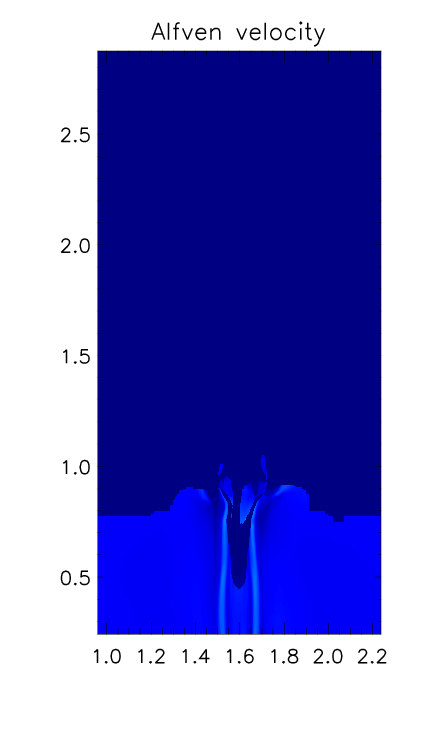}
\includegraphics[trim=0 50 0 50,clip, width=0.304\textwidth]{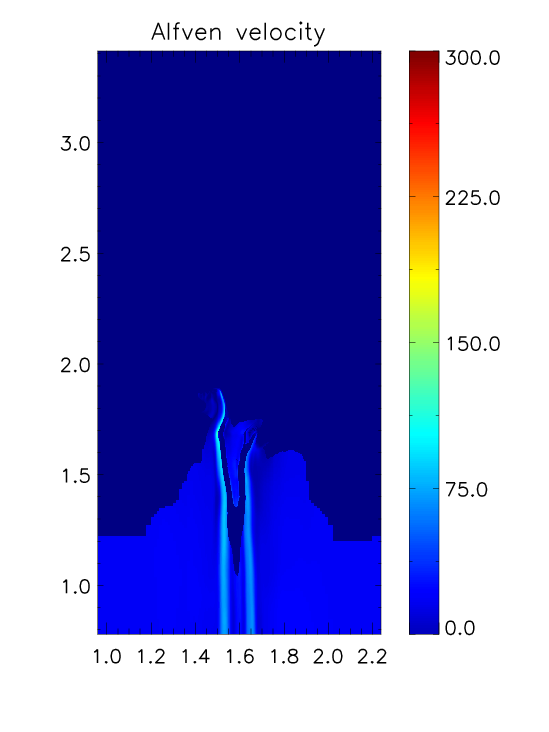}\\
\includegraphics[trim=0 50 0 50,clip, width=0.24\textwidth]{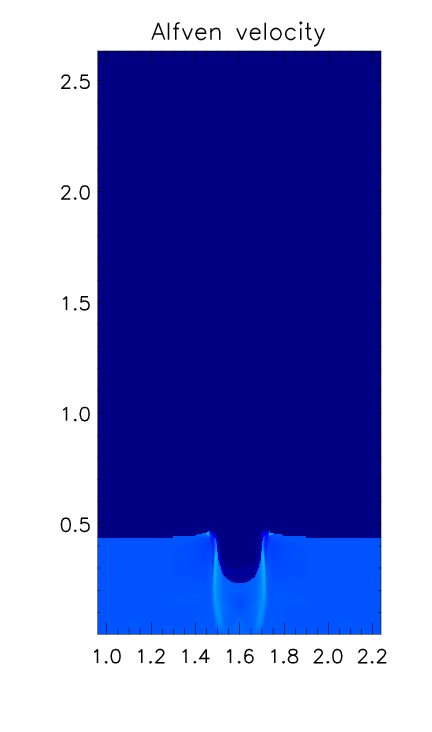}
\includegraphics[trim=0 50 0 50,clip, width=0.24\textwidth]{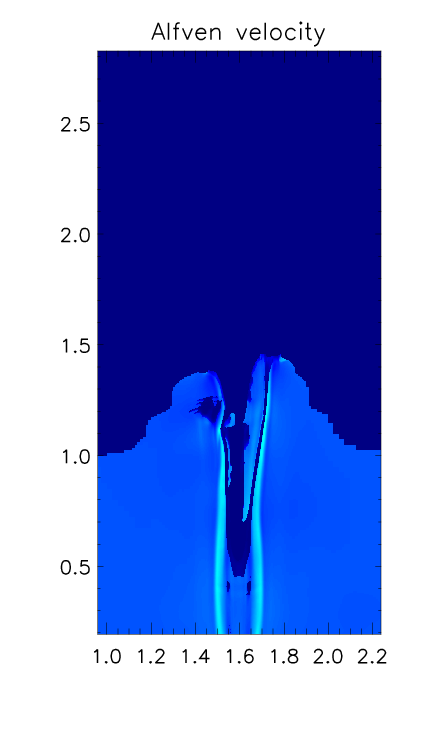}
\includegraphics[trim=0 50 0 50,clip, width=0.304\textwidth]{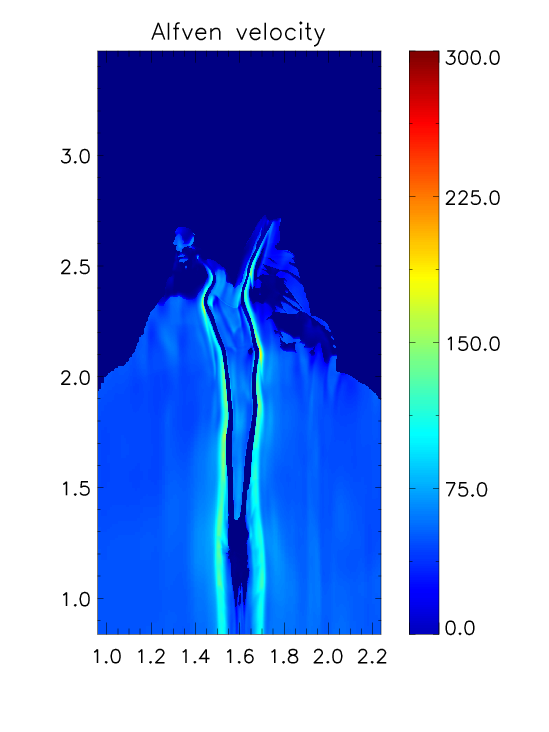}
\caption{Snapshots of Alfv\'en velocity at   $t=1$ (left), 5 (center),  and 10 $t_{\rm cc}$ (right). From top to bottom: CHI30BETA10, CHI30BETA3, CHI30BETA1. .}
\label{ax30}
\end{center}
\end{figure*}

 \begin{figure*}[htbp]
\begin{center}
\includegraphics[trim=0 50 0 50,clip, width=0.24\textwidth]{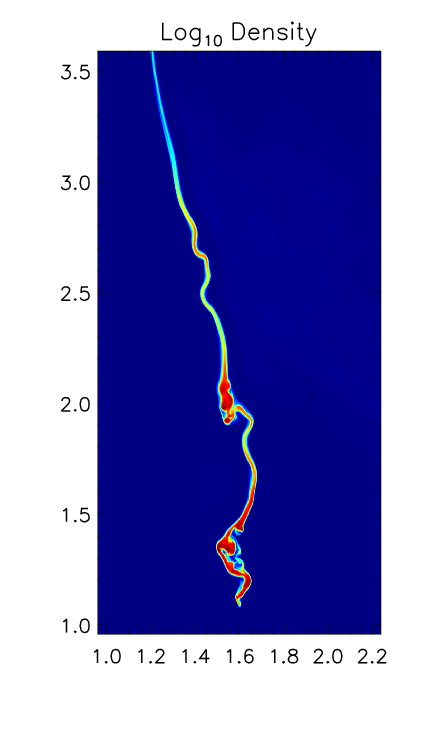}
\includegraphics[trim=0 50 0 50,clip, width=0.24\textwidth]{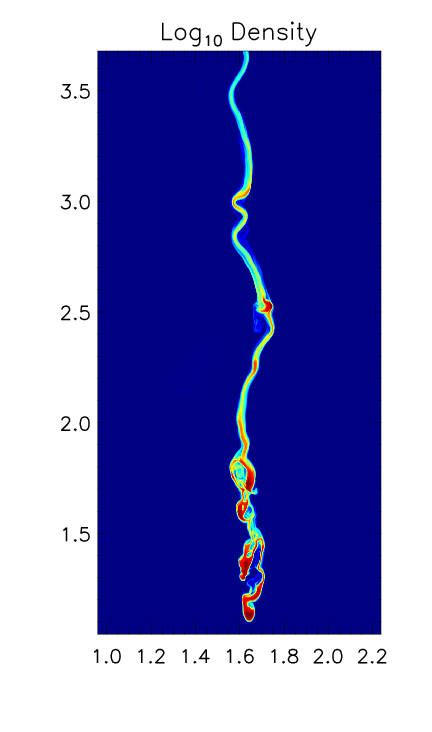}
\includegraphics[trim=0 50 0 50,clip, width=0.304\textwidth]{CHI300BETA3_COOL_200_cb_dens.png}\\
\includegraphics[trim=0 50 0 50,clip, width=0.24\textwidth]{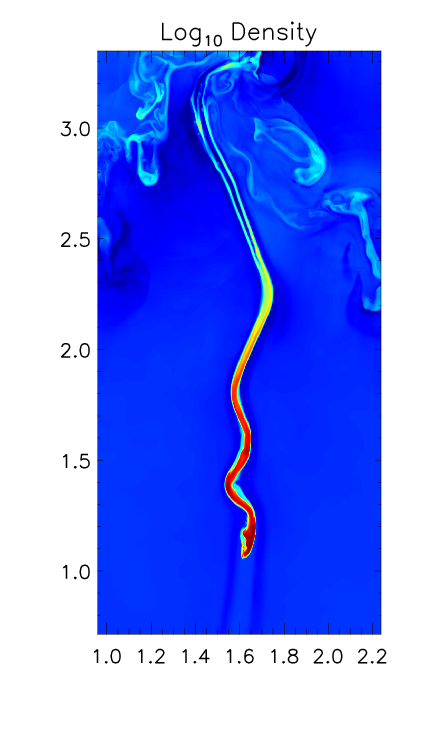}
\includegraphics[trim=0 50 0 50,clip, width=0.24\textwidth]{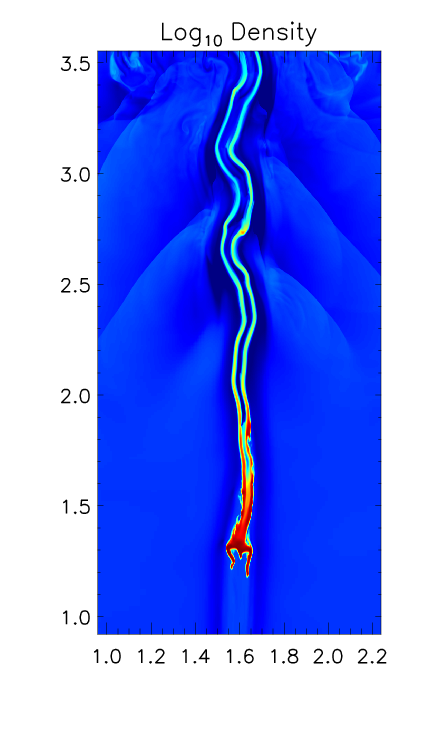}
\includegraphics[trim=0 50 0 50,clip, width=0.304\textwidth]{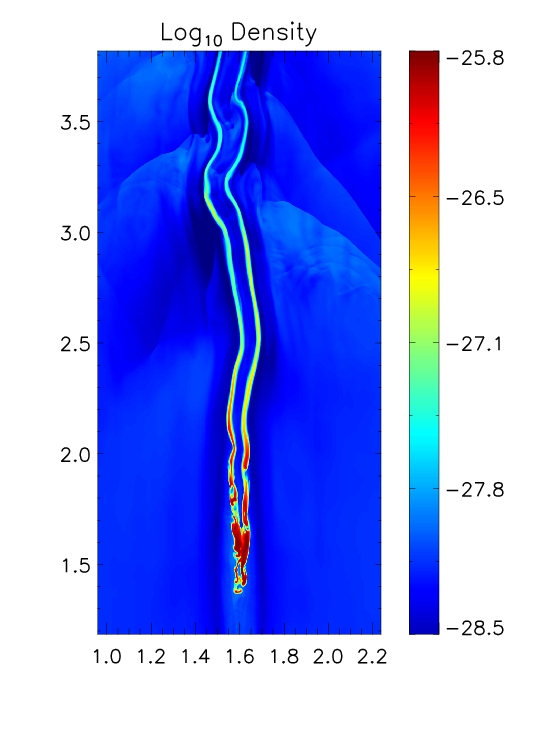}\\
\includegraphics[trim=0 50 0 50,clip, width=0.24\textwidth]{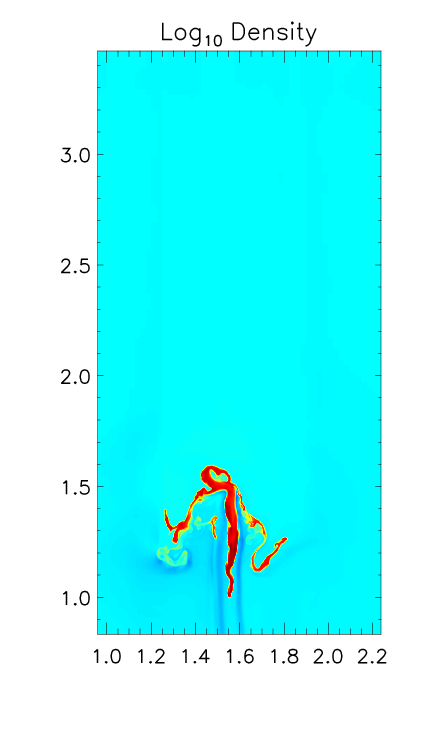}
\includegraphics[trim=0 50 0 50,clip, width=0.24\textwidth]{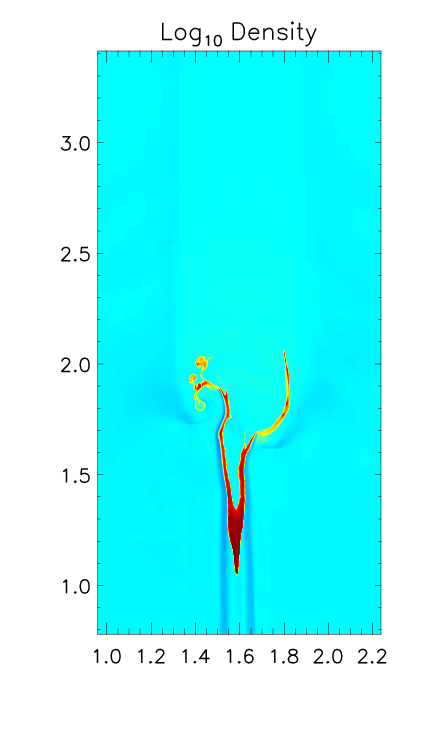}
\includegraphics[trim=0 50 0 50,clip, width=0.304\textwidth]{CHI30BETA1_COOL_200_cb_dens.png}
\caption{Snapshots of gas density at  $t= 10 t_{\rm cc}$.  From  left to right the columns represent runs with $\beta=100$, $10$ and $3$, respectively, while from top to bottom, the rows represent runs with $\chi=300,$ 100 and 30.}
\label{dens10tcc}
\end{center}
\end{figure*}

Figs.~\ref{dens30} and \ref{ax30} show the evolution of the cloud density and Alfv\'en speed in the $\chi=30$ case, again at three representative times of 1, 5,  and 10 $t_{\rm cc}.$   Here the same overall evolution occurs as is seen in the $\chi=300$ runs, with a bottleneck forming in front of the cloud, followed by the development of a boundary layer with a high Alfv\'en speed immediately surrounding the cloud.   Also as in the $\chi=300$ case, the bottleneck region is primarily responsible for cloud acceleration, while the boundary layer is responsible for most of the shear that disrupts the cloud.  As the density of the medium surrounding the cloud is 10 times higher than in the previous case, and hence the Alfv\'en speed is smaller, the filaments in this case are less extended than their $\chi=300$ counterparts. 

Finally, Fig.~\ref{dens10tcc} contrasts the density evolution of runs as a function of $\beta$ and $\chi.$  In general, the differences between the runs are primarily seen in the length of the filaments. These are most extended in those runs with the highest Alfv\'en speeds, which is $\propto \beta^{-1/2} \chi^{1/2}$ since the cloud density is the same in all cases.  On the other hand, as the CR pressure is similar in all the runs, the densest regions near the front of the cloud evolve similarly, regardless of $\beta$ and $\chi$.

It is important to point out that our results differ in a several significant respects from those in \cite{Wiener2019}.  As they do not adopt the techniques described in 2.1, particularly the regularization criterion, small waves develop in the ambient medium, which impede the streaming of the CRs past the cloud,  This  leads to an acceleration of the gas that surrounds the dense cloud and causes a bulk flow that contributes to the acceleration of the cloud and its subsequent evolution.  However, in simulations, this acceleration is dependent on numerics and, in nature, it is likely subdominant to acceleration of the hot material as it expands as  free wind.  As this bulk flow does not exist in our case, and instead the cloud evolution is driven by the CR pressure that builds up in front of the cloud and the shear layer that forms around the cloud, which leads to the onset of Kelvin-Helmholtz instabilities and the loss of mass. It is not obvious which computational set-up is more realistic.  In reality, CRs may well lead to the bulk acceleration of gas, but this may also not be the case,  so our focus is to fully understand the evolution caused purely by the CRs themselves.

\subsection{Mass Evolution}

\begin{figure}[htbp]
\begin{center}
\includegraphics[trim=0 0 20 0, clip,width=0.5\textwidth]{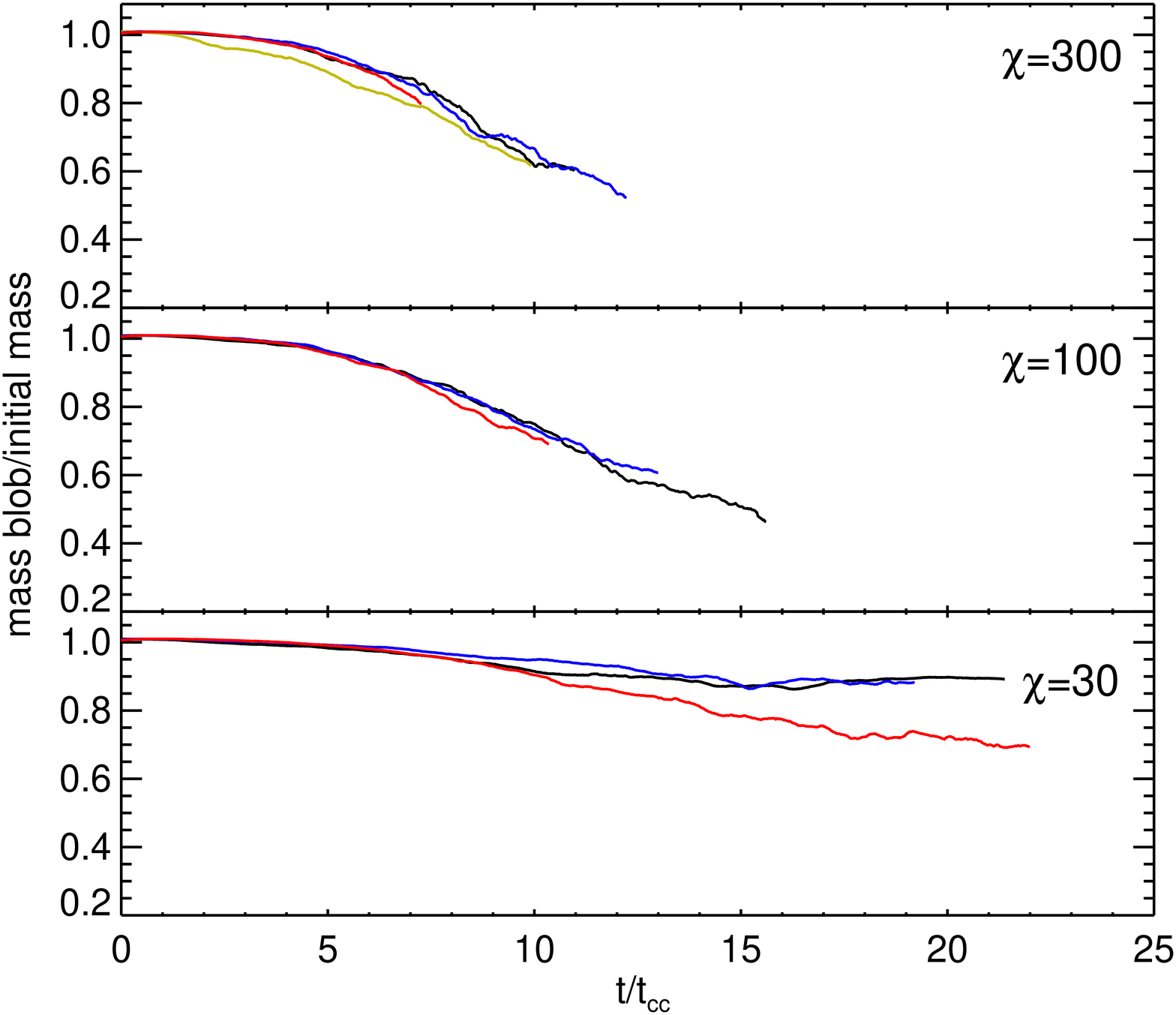}
\caption{Mass evolution for $\chi=300$ (top), $\chi=100$ (middle), $\chi=30$ (bottom) as a function of the cloud crushing time. In each panel the different lines correspond to the different values of $\beta$: $\beta=100$ (yellow), $\beta=10$ (black), $\beta=3$ (blue), $\beta=1$ (red).}
\label{massplot}
\end{center}
\end{figure} 

In Fig.~\ref{massplot}, we show the fraction of the mass retained by the cloud in each of the runs as a function of time in units of the cloud crushing time, $t_{\rm cc}$. More specifically, we plot $F_{1/3}(t)$,  which is the fraction of the total mass at or above 1/3 the original cloud density as defined in paper I.   Note that in this figure and the ones below, we define $t=0$ at two Alfv\'en times from the start of the simulation, as this is the time that the CRs first encounter the front of the cloud.

As the cloud crushing time only depends on $\chi$, the quantity $t/t_{\rm cc}$ is the same for all plots in the same panel. Hence, we find that the mass evolution at a given $\chi$ value is largely independent of $\beta$, and thus also largely independent of the Alfv\'en speed. The reason is that is that at a fixed pressure (CR + thermal), objects with the same size feel the same force, and hence the acceleration is proportional to the inverse of the mass. The higher Alfv\'en speed leads to a faster propagation of the stripped-off filaments, but since their density remains high, this does not translate into significantly faster mass loss. In every case, the Alfv\'en speed is sufficient to ensure that the CR pressure at the front of the bubble is roughly the same as at the inner $x$-boundary. This was verified by tracking the pressure at the bow of the cloud.

Comparing between runs with different $\chi$ values, we find that when plotted in $t_{\rm cc}$ units, the evolution is similar across models.  This means that that cloud disruption timescale goes roughly as $\chi^{1/2},$ as expected in the case of KH-driven mass loss \cite{1961hhs..book.....C,1974JFM....64..775B}.  However, in the low $\chi$ case, the evolution deviates somewhat from this trend, with the clouds surviving longest in the cases with the largest $\beta$s.  These are the cases with the lowest Alfv\'en speeds, in which the CR propagate most slowly, and the filaments are the least extended.  This delay in the development of an extended shear layer slows the evolution of the KH instability somewhat, preserving the cloud for more cloud crushing times.

\subsection{Velocity Evolution}

\begin{figure}[htbp]
\begin{center}
\includegraphics[trim=0 0 20 0, clip, width=0.5\textwidth]{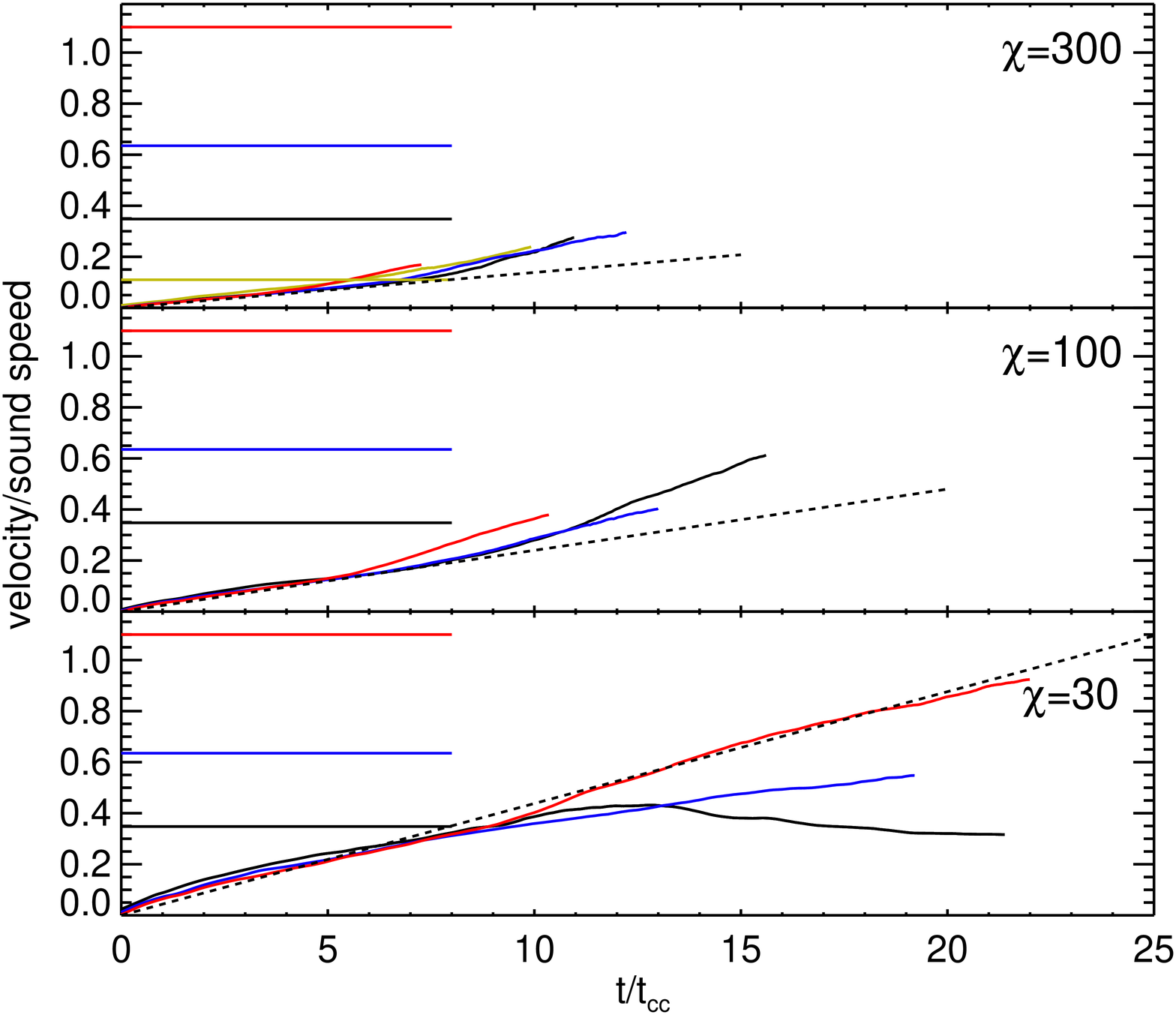}
\caption{Velocity (in $x$-direction) for  $\chi=300$ (top), $\chi=100$ (middle), $\chi=30$ (bottom) as a function of the cloud crushing time. In each panel the different lines correspond to the Alfv\'en speeds corresponding to the different values of $\beta$: $\beta=10$ (black), $\beta=3$ (blue), $\beta=1$ (red). The horizontal line shows the values of the Alfv\'en speeds in the respective runs, which the dashed line shows the evolution predicted by Eq.~(\ref{v_predict}) for $f=0.2$.}
\label{velplot}
\end{center}
\end{figure}

In Fig.~\ref{velplot}, we show the cloud velocity in terms of the external sound speed as a function of time in units of the cloud crushing time. As a simple model to guide the interpretation of these results, we consider the case in which the cloud cross section is fixed and the pressure on the cloud is constant and equal to the CR pressure at the boundary. In our two-dimensional configuration, this gives an estimate of 
\be
v_{0,\rm cl} = f \,   \frac{2 p_{\rm CR}}{R_{\rm cl} \rho_{\rm cl}}  \, t ,
\label{eq:v}
\ee
where $f$ is a factor that accounts for the fact that the cloud cross section drops as it is compressed and stretched during the interaction. Writing $t$ in units of the cloud crushing time and $v$ in units of the exterior sound speed, $c_s$, gives
\be
v(t)/c_s = f  \, \frac{2 p_{\rm CR} \chi^{1/2} }{\rho_{\rm cl} c_s^2}   \, \frac{t}{t_{\rm cc}}  =  f   \, 1.2  \chi^{-1/2} \, \frac{t}{t_{\rm cc}} ,
 \label{v_predict}
\ee
where in the second equality we have taken 
 $c_s $= 15 km s$^{-1}$ $\chi^{1/2},$ $p_{\rm CR}  = 1.4 \times 10^{-14}$ ergs cm$^{-3},$ and $\rho_{\rm cl}= 10^{-26}$ g cm$^{-3}$ as in our simulations.

In Fig.~\ref{velplot}, we overlay this prediction of Eq.~(\ref{v_predict}) for $f=0.2$ with the results of our simulations.  As in Fig.~\ref{massplot} we plot the velocity only for the material  at or above 1/3 the original cloud density. Here we see that at early times, this simple model is a good fit to the data, with $f=0.2$ accounting for the reduced cross-section of the cloud cause by the lateral squeezing by the CRs. At later times, however, notable deviations arise.  In the low $\beta$ high $\chi$ cases, the late time $x$-velocities exceed our simple estimate.  These are the cases in which the Alfv\'en speeds are the largest, and a significant fraction of the cloud mass is contained in the filaments, which accelerate faster than the head of the cloud.  In the high $\beta$ low $\chi$ cases, on the other hand, the late time $x$-velocities are roughly constant.  These are the cases in which the clouds start to approach the Alfv\'en speed, meaning that they are traveling at a rate similar to that of the CRs.  This reduces the pressure due to the bottleneck effect, leading to significantly lower $x$-velocities than estimated in Eq.~(\ref{v_predict}).

This general scaling is similar to what is observed in  \cite{Wiener2019}, in which the velocity of the cloud at a fixed time was found to be proportional to the CR energy density.  However, the set-up and the driving mechanisms are very different between these simulations.  While we hold the energy density of the CRs at the leftmost boundary constant, allowing them to stream onto the grid at the Alfv\'en speed,  \cite{Wiener2019} deposit CRs at a pre-defined rate in a fixed portion of the grid.  In addition, the acceleration of the cloud is both due the bottleneck effect studied here, as well as acceleration of the low density medium,  which does not arise in our work due to the numerical choices described above.  Finally, unlike  \cite{Wiener2019}, we do not switch off the CR pressure after a finite time, and thus we achieve velocities of the order of the Alfv\'en velocities even for smaller CR fluxes than those they investigated. Nevertheless, in both cases the velocity of the cloud at a fixed time is proportional to $p_{\rm CR}$ as arises in pressure estimate as in Eq.~(\ref{eq:v}), rather than proportional to $p_{\rm CR}^{1/2}$ as would occur in a situation in which a fixed fraction of the CR energy density is being transferred to the cloud.

\subsection{The Effect of Radiative Cooling}

\begin{figure*}[htbp]
\begin{center}
\includegraphics[trim=0 50 0 50,clip, width=0.22\textwidth]{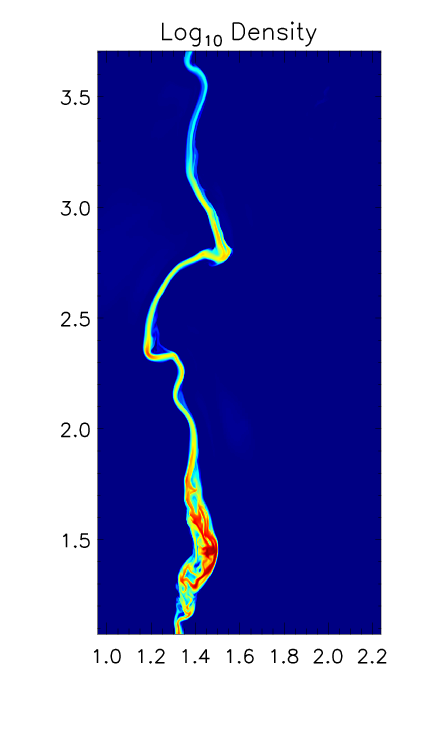}
\includegraphics[trim=0 50 0 50,clip, width=0.22\textwidth]{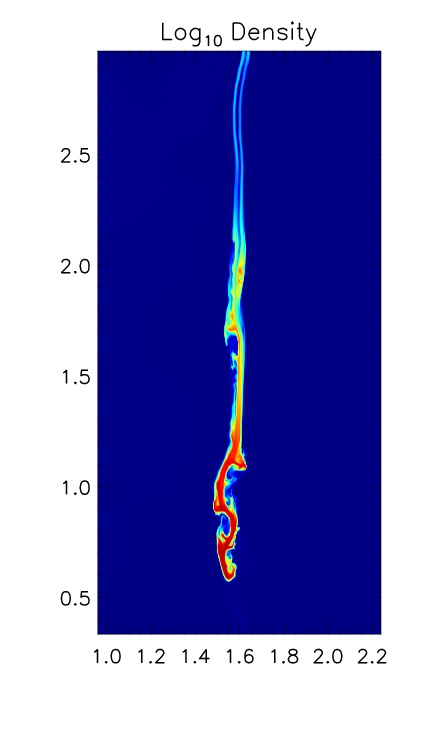}
\includegraphics[trim=0 50 0 50,clip, width=0.278\textwidth]{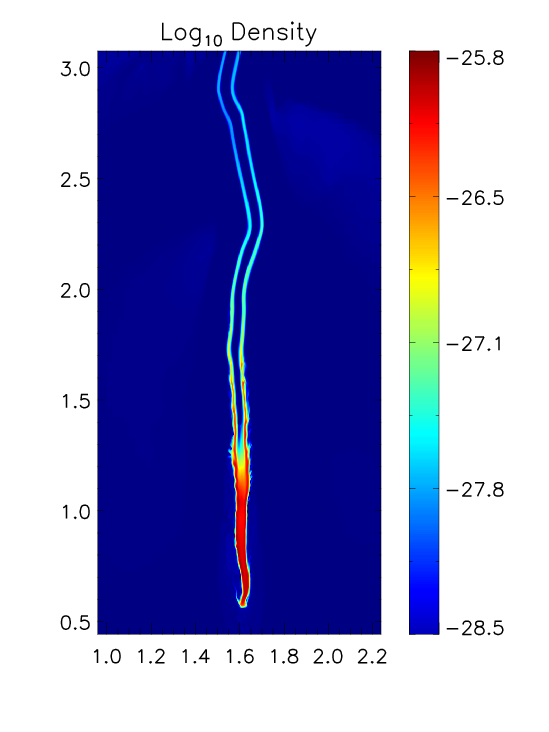}\\
\includegraphics[trim=0 50 0 50,clip, width=0.22\textwidth]{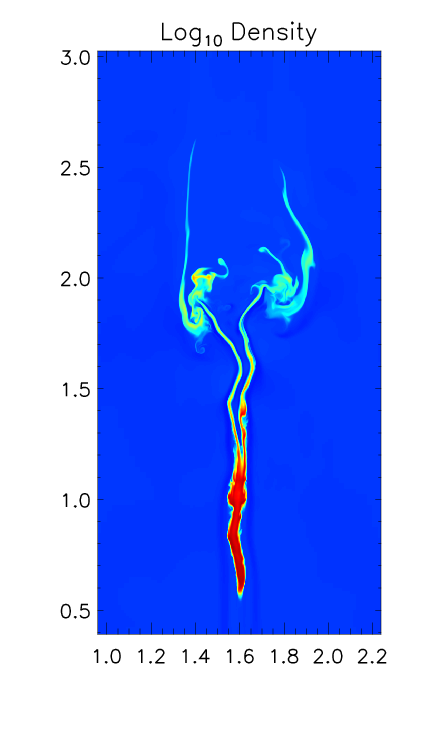}
\includegraphics[trim=0 50 0 50,clip, width=0.24\textwidth]{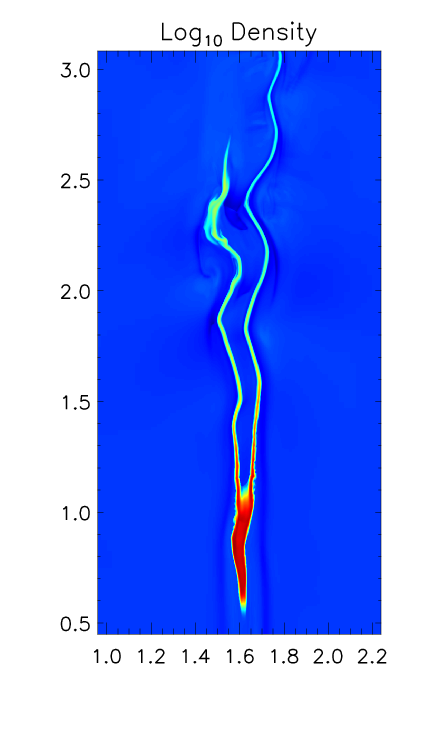}
\includegraphics[trim=0 50 0 50,clip, width=0.278\textwidth]{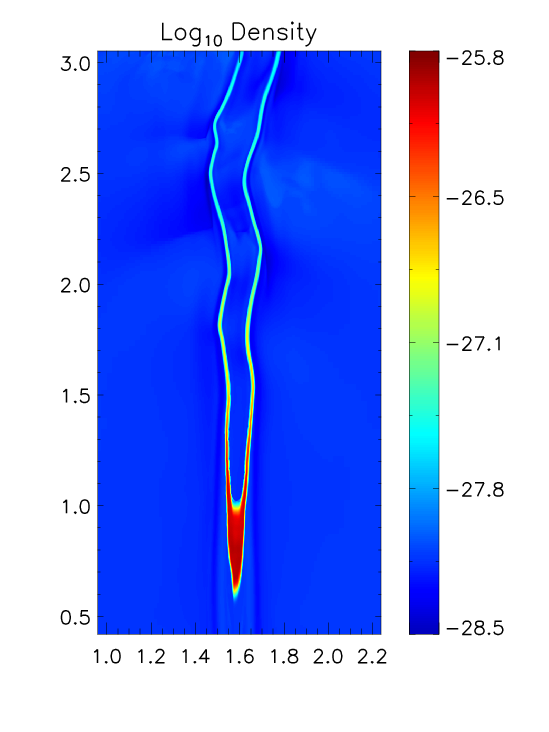}\\
\includegraphics[trim=0 50 0 50,clip, width=0.22\textwidth]{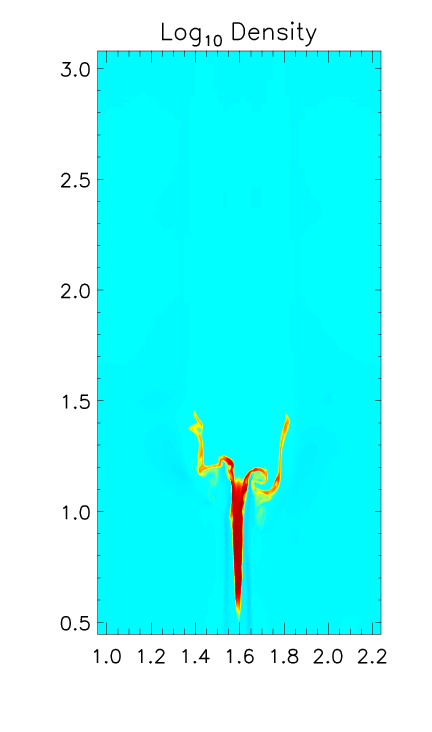}
\includegraphics[trim=0 50 0 50,clip, width=0.22\textwidth]{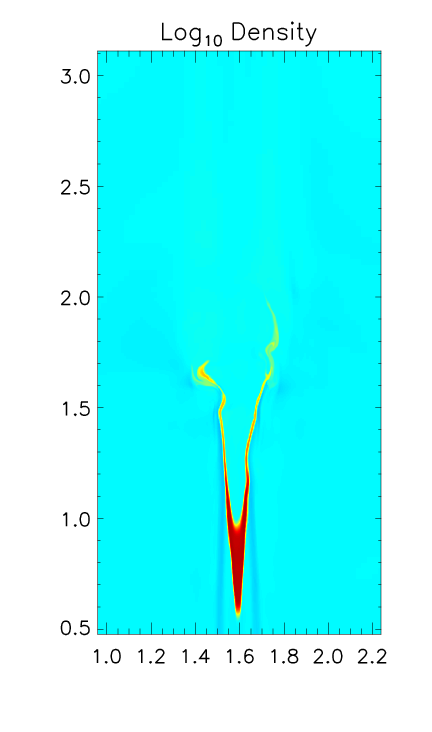}
\includegraphics[trim=0 50 0 50,clip, width=0.278\textwidth]{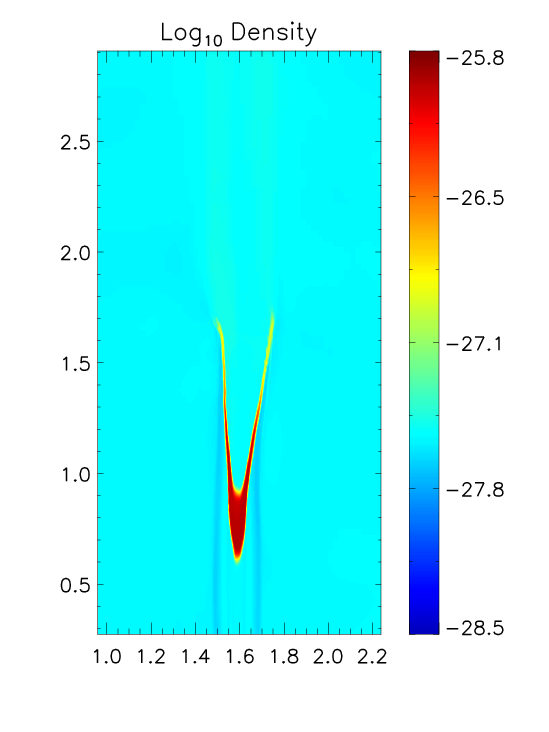}
\caption{Snapshots of gas density at  $t = 10 t_{\rm cc}$ with no radiative cooling.  Snapshots of gas density at  $t= 10 t_{\rm cc}$.  From  left to right the columns represent runs with $\beta=100$, $10$ and $3$, respectively, while from top to bottom, the rows represent runs with $\chi=300,$ 100 and 30. The axes labels denote kpc.}
\label{dens5tcc_nocool}
\end{center}
\end{figure*}

In order to assess the effect of cooling, we ran a selection of runs with radiative cooling switched off.   Radiative cooling is known to delay mixing by keeping the boundary layer thinner and by preventing the growth of the KH instability \citep[e.g.][]{1997ApJ...483..262V}. However, once the KH instability turns nonlinear, it is not clear how effectively cooling can suppress mixing \citep{2000A&A...360..795M}. 

 In Fig.~\ref{dens5tcc_nocool} we show the gas density at  $=10 t_{\rm cc},$ from nine such simulations. Unlike ram-press acceleration, acceleration by the CRs does not produce shocks that heat the cloud, and must be radiated away to avoid disruption.  In fact, we find that switching off radiative cooling does not lead to qualitatively different behavior. In this case, the filaments that are smoother than in the previous runs, suggesting that cooling causes more fragmentation at the shear interface between the cloud and the ambient medium.  However, the overall morphology of the clouds are similar, which also translates into a similar evolution of the mass and the velocity of the cloud. This is in agreement with \cite{Wiener2019} (see their Fig. 14 that, however, still includes CR heating).

\begin{figure}[htbp]
\begin{center}
\includegraphics[trim=0 0 20 0, clip,width=0.5\textwidth]{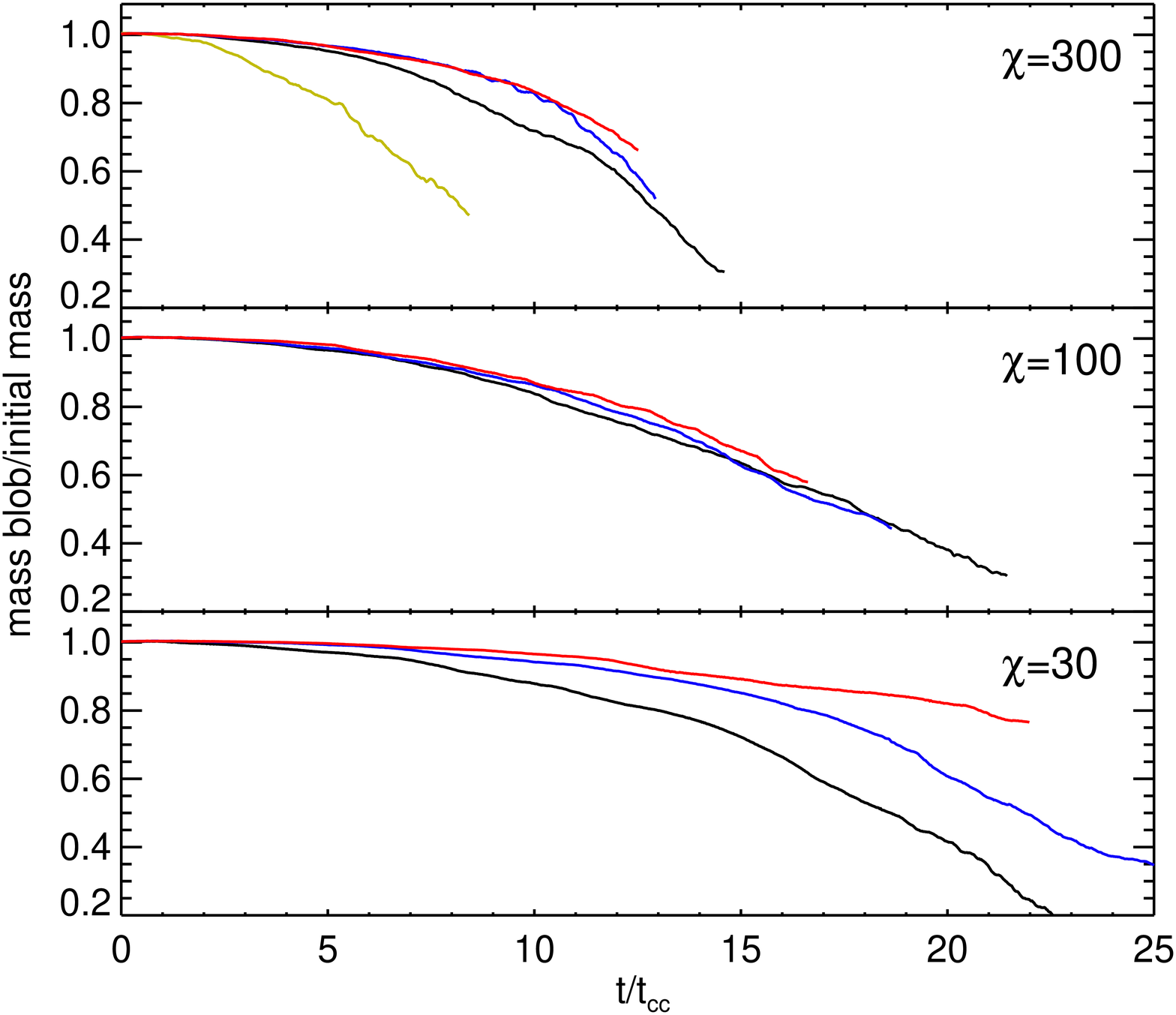}
\caption{Mass evolution for $\chi=300$ (top), $\chi=100$ (middle), $\chi=30$ (bottom) as a function of the cloud crushing time in the absence of radiative cooling. In each panel the different lines correspond to the different values of $\beta$: $\beta=100$ (yellow), $\beta=10$ (black), $\beta=3$ (blue), $\beta=1$ (red).}
\label{massplotnc}
\end{center}
\end{figure}

 In Fig.~\ref{massplotnc}, we show the fraction of the mass retained by the cloud in each of these runs as a function of time in units of the cloud crushing time, $t_{\rm cc}$. Again, this figure is very similar to Fig.~\ref{massplot}. There are some differences, most notably for $\chi=30$, where the mass declines more quickly than in the case with radiative cooling. Also, for $\chi=300$ and $\beta=100$ the mass of the cloud declines more quickly than  in the case with radiative cooling. In this run, which is closest to the pure hydrodynamics case and radiative cooling is most needed to keep the cloud intact and impede the ablation of gas. For the cases with lower $\beta$ this is less pronounced because the magnetic field acts to keep the cloud from disrupting.

\cite{Wiener2019} stress that the inclusion of radiative cooling prevents the disruption of the cloud by CR heating. Such heating leads to an additional term, ${\cal H}_{\rm CR} = - v_{\rm A} \cdot \nabla p_{\rm CR}$ \citep{1971ApJ...163..503W}, that goes into Eq.~(8) but is not included in the work presented here. However, the disruption time scale via cosmic-ray heating for our choice of parameters is of the order of 100 Myrs, which is too long to have an effect on cloud evolution on the time scales relevant for our simulations. 

Finally, to assess the impact of the numerical switch that allows acceleration by CRs only on cells whose density lies at least 10 percent above the ambient density, we carried out a comparison run for the $\chi=100$, $\beta=10$ case in which the switch was turned off, but the regularization was maintained.  In this run, the mass loss rate and acceleration of the blob are very similar to the fiducial case at early times ($t \leq 5 t_{\rm cc}$), and slightly greater (by about a factor of $1.5$) at later times.  These differences are small enough such that they do not affect any of the conclusions described above.  Furthermore, as they are caused by CRs scattering off of numerical noise, they are likely to be resolution-dependent, and smaller in higher-resolution simulations.

\section{Conclusions}

The acceleration of cold clouds by cosmic rays is an important topic in understanding the multiphase evolution of starburst-driven galactic winds.  Such outflows contain not only hot $\approx 10^7-10^8$ K material, with sound speeds that are comparable to the outflow velocities, but also $\approx 10^4$K clouds, whose sound speeds are less than an order of magnitude smaller.  While this supersonic gas is often assumed to be accelerated by the hot wind, such interactions tend to shred and evaporate the clouds before significant acceleration is achieved.  At the same time, the supernovae that drive starbursts also generate a large number of cosmic rays, which can potentially accelerate cold gas while avoiding the destructive processes observed in ram-pressure acceleration.

To better understand this possibility, we have simulated the interactions of cold clouds with CR gradients of the type expected in galactic winds. Using the FLASH AMR MHD code, we exposed cold clouds to CR gradients that stream with Alfv\'en speeds along the magnetic field lines.  These 2D simulations include radiative cooling, but do not include explicit cosmic ray heating, and they include an important switch that suppresses the acceleration of the ambient medium by the streaming against sound waves and weak shocks.

In this case, the CRs stream towards the cloud at the Alfv\'en speed, which decreases dramatically at the cloud boundary, leading to a  bottleneck in which the CRs pile up in front of the cloud,  building up pressure. At the same time, the cosmic rays start to stream along the sides of the cloud, causing the cloud to be compressed in the perpendicular direction. The low densities and high field values in this boundary layer cause the Alfv\'en speed to go up dramatically, leading to significant shear.  By sampling a large range of the density contrasts, $\chi,$ and ratios of thermal to magnetic pressure, $\beta$, we are able to study how these effects lead to acceleration and mass loss over the conditions experienced by cold clouds.

The major results from these simulations are:

\begin{itemize}

\item Cloud destruction occurs primarily due to shear in the filaments that arise in the boundary layer.  As expected from the KH instability, this leads to a mass loss rate in units of $t_{\rm cc}$ that is largely independent of both $\chi$ and $\beta$.  In most cases, the clouds lose half their mass by $\approx 12 t_{\rm cc}$, although the mass loss rate is slower in the cases with very low $\chi$ and $\beta$. These are the cases with the lowest Alfv\'en speeds, in which the CRs take the most time to propagate, the filaments are the least extended, and the KH shear layer develops most slowly.

\item The acceleration of the cloud is consistent with a simple model in which CR pressure is constant and the cloud cross section is $\approx 0.2$ of its initial cloud size.  This reduction in the size of the cloud perpendicular to the flow is caused by the pressure of the CRs streaming around the cloud, and it leads to a cloud velocity of $\approx 0.25 c_s \chi^{-1/2} t/t_{\rm cc}.$  In most cases, this is long enough to allow for cloud acceleration to a significant fraction of the exterior sound speed. However, in cases in which clouds approach the Alfv\'en speed, they begin to outrun the impinging CRs, leading to more gradual acceleration.  

\item Radiative cooling has relatively little impact on cloud evolution. By carrying out comparison runs in which it was switched off,  we found that the overall cloud morphology is somewhat smoother than in runs with cooling.  However, mass loss rates and accelerations were similar in both cases.

\item Together, these results show that CRs, if acting as the primary sources of momentum input, are capable of accelerating clouds to velocities comparable to those observed in galaxy outflows. 

\end{itemize}

There are, however, a number of important caveats to this conclusion. Our treatment of CR transport does not consider the effects of wave damping, which may become relevant inside the cold cloud if Alfv\'en waves suffer significant ion-neutral damping. This could increase the CR energy density inside the cloud at the beginning of the interaction.  Our results also do not include CR heating, which can have a significant impact for more extreme CR pressure gradients and on longer time scales than the ones simulated here. Our simulations assume that magnetic fields penetrate the clouds, which is a necessary condition to produce CR  bottlenecks. Finally, our simulations are two-dimensional, due to restrictive timestep requirements, which may alleviated in the future by a new method to compute CR streaming that is based on a radiative transfer method \citep{2018ApJ...854....5J,  2019MNRAS.485.2977T}.

A key difference between our work and the work described in  \cite{Wiener2019} is due to our choice of regularization criteria and the threshold on the CR gradient in the $x$-direction. Together, these prevent the ambient medium from being significantly accelerated by CRs, in contrast to the significant acceleration of the ambient medium seen in this previous work.  The acceleration of the exterior medium causes substantial shear at the cloud-environment interface, influencing the cloud evolution in ways that are distinct and difficult to disentangle from effects due to the CR pressure.

While the CR acceleration of the exterior medium remains an open question, it will be subdominant with respect to the acceleration of the ambient medium due its thermal pressure.   This is because,  unlike cold clouds, the hot medium has sound speed much higher than the escape velocity of the host galaxy, and it can expand freely into the surrounding environment. This means that cold cloud evolution in galactic outflows will likely depend on the combination of CRs and gas shear that is dependent on the particulars of the starburst itself and the magnetic field structure near the clouds, and that cloud survival will depend on the evolution of the KH instability in this complex situation.  A better understanding of such interactions will continue to improve our understanding of the cold clouds observed in starburst-driven galactic winds.

\acknowledgements
We would like to thank the reviewer for a very constructive review. Part of this work was performed at the Turbulent Life of Cosmic Baryons program at the Aspen Center for Physics, which is supported by National Science Foundation grant PHY-1607611. We thank Josh Wiener and Ellen Zweibel for helpful discussions. The authors gratefully acknowledge the Gauss Centre for Supercomputing e.V. (www.gauss-centre.eu) for funding this project (16072) by providing computing time through the John von Neumann Institute for Computing (NIC) on the GCS Supercomputer JUWELS at J\"ulich Supercomputing Centre (JSC). The FLASH code was developed in part by the DOE-supported  Alliances Center for Astrophysical Thermonuclear Flashes (ASC) at the University of Chicago. ES was supported by NSF grant AST-1715876.

\bibliographystyle{apj}
\bibliography{CosmicRay}

\end{document}